\documentclass[aps,twocolumn,showlabels,showrefs,amsmath,amssymb,pre,superscriptaddress,floatfix,colors]{revtex4-2}

\usepackage{array}
\usepackage{graphicx,epsfig}
\usepackage{subfigure} 
\usepackage{droidsans}
\usepackage{charter}
\usepackage[T1]{fontenc}
\usepackage[usenames,dvipsnames]{xcolor}
\usepackage{setspace}
\usepackage[compact]{titlesec}
\usepackage{hyperref}
 \usepackage{amsmath,amssymb}
\usepackage{bm}% bold math
\usepackage{fancyhdr}
\usepackage[capitalise]{cleveref}
\usepackage{bbm}

\newcommand{\Da}{\mathcal{D}}

\begin{document}

\title{Impact of dipole-dipole interactions on motility-induced phase separation}

\author{Elena Ses\'{e}-Sansa}
\affiliation{CECAM,  Centre  Europ\'een  de  Calcul  Atomique  et  Mol\'eculaire, \'Ecole  Polytechnique  F\'ed\'erale  de  Lausanne (EPFL),  Batochime,  Avenue  Forel  2,  1015  Lausanne}\email{elena.sesesansa@epfl.ch} %optional
\author{Guo-Jun Liao}
\affiliation{Institut f\"{u}r Theoretische Physik, Technische Universit\"{a}t Berlin, Hardenbergstr. 36, D-10623 Berlin, Germany} %optional
\author{Demian~Levis} 
\affiliation{Departament de F\'isica de la Mat\`eria Condensada, Universitat de Barcelona, Mart\'{\i} i Franqu\`es 1, E08028 Barcelona, Spain}
\affiliation{UBICS  University  of  Barcelona  Institute  of  Complex  Systems, Mart\'{\i} i Franqu\`es  1, E08028 Barcelona, Spain}
\author{Ignacio~Pagonabarraga} 
\affiliation{CECAM,  Centre  Europ\'een  de  Calcul  Atomique  et  Mol\'eculaire, \'Ecole  Polytechnique  F\'ed\'erale  de  Lausanne (EPFL),  Batochime,  Avenue  Forel  2,  1015  Lausanne}
\affiliation{Departament de F\'isica de la Mat\`eria Condensada, Universitat de Barcelona, Mart\'{\i} i Franqu\`es 1, E08028 Barcelona, Spain}
\affiliation{UBICS  University  of  Barcelona  Institute  of  Complex  Systems, Mart\'{\i} i Franqu\`es  1, E08028 Barcelona, Spain}
\author{Sabine H. L. Klapp}
\affiliation{Institut f\"{u}r Theoretische Physik, Technische Universit\"{a}t Berlin, Hardenbergstr. 36, D-10623 Berlin, Germany} %optional

\begin{abstract}
We present a hydrodynamic theory for systems of dipolar active Brownian particles which, in the regime of weak dipolar coupling, predicts the onset of motility-induced phase separation (MIPS), consistent with Brownian dynamics (BD) simulations. The hydrodynamic equations are derived by explicitly coarse-graining the microscopic Langevin dynamics, thus allowing for a quantitative comparison of parameters entering the coarse-grained model and particle-resolved simulations. Performing BD simulations at fixed density, we find that dipolar interactions tend to hinder MIPS, as first reported in \ [Liao \textit{et al., Soft Matter}, 2020, \textbf{16}, 2208]. Here we demonstrate that the theoretical approach indeed captures the suppression of MIPS. Moreover, the analysis of the numerically obtained, angle-dependent correlation functions sheds light into the underlying microscopic mechanisms leading to the destabilization of the homogeneous phase.
\end{abstract}

\maketitle

%Please use \dag to cite the ESI in the main text of the article.
%If you article does not have ESI please remove the the \dag symbol from the title and the footnotetext below.
%\footnotetext{\dag~Electronic Supplementary Information (ESI) available: [details of any supplementary information available should be included here]. See DOI: 10.1039/cXsm00000x/}
%additional addresses can be cited as above using the lower-case letters, c, d, e... If all authors are from the same address, no letter is required

%\footnotetext{\ddag~Additional footnotes to the title and authors can be included \textit{e.g.}\ `Present address:' or `These authors contributed equally to this work' as above using the symbols: \ddag, \textsection, and \P. Please place the appropriate symbol next to the author's name and include a \texttt{\textbackslash footnotetext} entry in the the correct place in the list.}

\section{Introduction}

Active systems are composed of large numbers of individual units that constantly consume energy at the level of each constituent and convert it into directed motion \cite{Ramaswamy2010,Bechinger2016}. Many examples on different length scales can be found in the biological world, ranging from animal groups \cite{Cavagna2010,Katz2011}, to cells \cite{Kessler1986,Saw2017} and bacteria \cite{Sokolov2007,Beer2019,klumpp2019,frankelmagnetic1984}. The increasing interest in these systems, which are intrinsically out of equilibrium, has also fostered the engineering of synthetic active particles designed and controlled in the lab. Prominent examples are active colloids capable of self-propulsion due to applied electric \cite{bricard2013,Yan2016,Yang2019,Sprenger2020,Fernandez-Rodriguez2020} or magnetic \cite{Snezhko2011,Lumay2013,Grosjean2015,Kaiser2017} fields, light \cite{Buttinoni2012, Palazzi2013} or chemical gradients \cite{Theurkauff2012,Buttinoni2013,Dietrich2017}. 

It is now well established that large ensembles of active particles display a wide variety of complex behaviors, like flocking \cite{Cavagna2010,Bhattacharya2010,Bialek2012}, swarming \cite{buhl2006,Beer2019,VanDerVaart2019}, mesoscale turbulence \cite{wensink2012,Kokot2017,Doostmohammadi2017,Reinken2018,Reinken2019} or laning \cite{Vissers2011,kogler2015,wachtler2016}. A further intriguing phenomenon taking place in active systems with steric interactions is 
a full phase separation leading to the coexistence of a dense cluster and a dilute gas-like phase. This phenomenon, known as \emph{motility-induced phase separation} (MIPS) \cite{Redner2013,stenhammar2014,Fily2012,Cates2015,solon2015,Digregorio2018}, somewhat resembles the vapor-liquid phase transition in equilibrium systems with attractive interactions. In active systems, however, the transition is solely induced by activity. In particular, MIPS results from the interplay between self-propulsion and steric effects, which triggers the slowing-down of particles upon collision and creates a positive feedback mechanism by which particles accumulate in dense regions and further slow down. 
MIPS has been reported in many numerical simulation studies of different models. These include, in particular, active Brownian particles (ABP) \cite{Fily2012,Romanczuk2012,Cates_runtumble2013,stenhammar2014,Digregorio2018}, arguably one of the simplest microscopic models of active matter, which describes isotropic self-propelled disks or spheres (2D or 3D models) interacting only through volume exclusion. At a theoretical level, MIPS has also been explained by means of various types of continuum theories \cite{Cates_runtumble2013,Bialke2013,Wittkowski2014,Speck2015,Nardini2017}. 

Recently, a line of research has specifically focused on understanding the impact of aligning mechanisms on the collective behavior of self-propelled particles \cite{Barre2015,Martin-Gomez2018,Shi2018,Sese-Sansa2018,Geyer2019,VanderLinden2019,VanDamme2019,Bhattacherjee2019,Caprini2020,frey2020,Bar2020,Grossmann2020,Jayaram2020,Liao2020,Zhang2021,Sese-Sansa2021}, paying special attention to the effect they have on MIPS. This has been tackled both in particle-resolved simulations \cite{Shi2018,Sese-Sansa2018,VanDamme2019,Caprini2020,Grossmann2020,Jayaram2020,Liao2020,Sese-Sansa2021} and also by deriving coarse-grained descriptions \cite{Geyer2019,Grossmann2020,Jayaram2020,Zhang2021,Sese-Sansa2021}, which often provide a deeper insight into the underlying mechanisms governing the system's phase behavior. It has been shown that the resulting impact of aligning interactions on MIPS depends very much on the origin and type of interactions, and resulting torques, considered. 
In the case of 'steric' alignment due to collisions of elongated particles, the phase separation induced by motility disappears as soon as shape anisotropy sets in \cite{Shi2018,VanDamme2019,Jayaram2020,Grossmann2020}. On the contrary, MIPS is promoted by torques leading to autotaxis \cite{Zhang2021} and also in the presence of polar velocity-alignment in discoidal self-propelled particles \cite{Barre2015,Sese-Sansa2018,Sese-Sansa2021}. Further, MIPS remains essentially unaffected by nematic interactions between disks \cite{Sese-Sansa2021}. 

In the present paper we consider the more complex case of dipolar interactions, which have been less investigated but appear in a number of experimental systems. Examples are ferromagnetic rollers which show flocking and vortex states in applied alternating fields \cite{Kaiser2017,han2020} or Janus colloids half-coated with a metallic cap \cite{Yan2016,Fernandez-Rodriguez2020,Zhang2021}. The resulting (induced) dipole-dipole interactions can give rise to swarming, chaining and clustered states.

It has been recently reported in numerical simulations of dipolar active Brownian particles that dipole-dipole interactions actually hinder MIPS \cite{Liao2020}. This finding is in contrast with, e.g., Vicsek-like interactions \cite{Barre2015,Sese-Sansa2018,Sese-Sansa2021}, showing that details of the anisotropy play an important role.
There is, however, so far no (coarse-grained) theory describing systems of self-propelled particles with dipolar interactions, which, due to their long-range nature and to the fact that they depend on both the orientation and spatial configuration of a pair of particles, pose a severe challenge for hydrodynamic theories. This is the goal of the present paper. 
%Such disparity of scenarios has led to the derivation of various coarse-grained descriptions to account for the effect that different types of aligning mechanisms have on MIPS.
We derive a continuum theory for systems of dipolar active disks starting from their microscopic dynamics, described by the $N$-body Smoluchowski equation. Specifically, we consider the model system first proposed in \cite{Liao2020}, which describes systems of ABP with an embedded point dipole moment in their center. In order to derive 
the theoretical coarse-grained model, we extend the approach introduced in \cite{Bialke2013,Speck2015} for pure ABP, to the case of dipole-dipole interactions. The theoretical framework allows for a direct mapping between the continuum and the microscopic model. We show that the hindering of MIPS due to dipolar interactions is indeed captured by the continuum description. Moreover, the coarse-grained model, combined with results of Brownian dynamics simulations, gives a quantitative prediction of the onset of phase separation. Therefore, our results help to elucidate the complex mechanisms behind the suppression of MIPS in the presence of dipolar interactions. 

The paper is organised as follows. In \cref{modelmethods} we introduce the microscopic model we consider. In \cref{theoreticaldesc} we present the derivation of the coarse-grained hydrodynamic model starting from an $N$-body Smoluchowski equation. We then perform a linear stability analysis that leads to the prediction of the onset of MIPS, in the presence of dipolar interactions. \cref{numeric_results} is devoted to the comparison between the microscopic model and the coarse-grained description. To this end, we first present results from Brownian dynamics simulations and we then explain how to establish the mapping between the particle-based model and the continuum description. This provides us with the tools to finally compare quantitatively both approaches.

\section{Model} \label{modelmethods}

We consider a two-dimensional system of $N$ disk-like particles with positions $\textbf{r}_i$ and orientations $\hat{\textbf{e}}_{i}=\left(\cos\varphi_i, \sin\varphi_i\right)^{T}$, $\varphi_i$ being the polar angle. The system's dynamics is governed by the overdamped Langevin equations

\begin{equation}
\dot{\textbf{r}}_{i} = \beta D_t \left(F_{0}\hat{\textbf{e}}_{i} - \nabla_i U+ \boldsymbol{\eta}_{i} \right),
\label{brown-dyn-mod_eq:1}
\end{equation}
\begin{equation}
\dot{\varphi}_{i} =  \beta D_r \left(-\partial_{\varphi_i} U+ \nu_{i}  \right).
\label{brown-dyn-mod_eq:2}
\end{equation}

The particles' positions are subject to thermal noise, $\boldsymbol{\eta}_{i}$, which mimics the impact of the surrounding solvent and has zero mean and delta-like temporal correlations, i. e., $\langle \boldsymbol{\eta}_{i}(t) \rangle = 0$, $\langle \boldsymbol{\eta}_{i}(t)\otimes\boldsymbol{\eta}_{j}(t')\rangle=2 \delta_{ij}\delta(t-t')\mathbb{I}/(\beta^2 D_t)$. Here, $\beta=(k_B T)^{-1}$ (with $k_B$ Boltzmann's constant and $T$ being the temperature) is the inverse thermal energy, $D_t$ is the translational diffusion coefficient and the symbol $\otimes$ represents the dyadic product. Likewise, orientations are subject to rotational noise, $\nu_{i} $, with $\langle \nu_{i}(t) \rangle = 0$, $\langle \nu_{i}(t),\nu_{j}(t')\rangle=2 \delta_{ij}\delta(t-t') /(\beta^2 D_r)$, and $D_r$ is the rotational diffusion coefficient.

Each particle self-propels at constant speed $v_0=\beta D_t F_0$ along the instantaneous orientation $\hat{\textbf{e}}_{i}$. In addition, particles carry a permanent dipole moment $\boldsymbol{\mu}_i = \mu \hat{\boldsymbol{\mu}}_i $ in their center. We choose $\hat{\boldsymbol{\mu}}_i = \hat{\textbf{e}}_{i}$, so that the particle's direction of self-propulsion always coincides with its dipole orientation.

%Particles interact through excluded-volume interactions, which we model using a WCA potential. In addition, we consider the interaction between dipole moments
 The conservative interaction potential between pairs of particles is assumed to have two contributions stemming from steric (excluded-volume) and dipole-dipole interactions,

\begin{equation}
\begin{split}
U&(\left\{ \textbf{r}_i \right\}, \left\{ \varphi_i \right\}) =\sum _{i=1}^{i=N} \Big( \sum_{i<j}^{j=N} u_{sr}(r_{ij} ) + \sum_{i<j}^{j=N} u_{dd}(\textbf{r}_{ij},\boldsymbol{\mu}_i,\boldsymbol{\mu}_j) \Big),
 \end{split}
\label{brown-dyn-mod_eq:3}
\end{equation}
where $r_{ij} = |\textbf{r}_{ij}|=|\textbf{r}_j - \textbf{r}_i |$.

The steric repulsion between the disks is modeled using a Weeks-Chandler-Anderson (WCA) potential,

\begin{equation}
\begin{split}
u_{sr}(r_{ij} ) =
\begin{cases} 
4u_0 \left(\left(\frac{\sigma}{r_{ij}}\right)^{12} - \left(\frac{\sigma}{r_{ij}}\right)^{6} + \frac{1}{4}  \right), & r_{ij} \le R\\
0, & r_{ij} > R,
\end{cases}
\end{split}
\label{brown-dyn-mod_eq:5}
\end{equation}
where the cutoff distance is $R = 2^{1/6} \sigma$, with $\sigma$ being the particle's diameter, and $ u_0$ the interaction strength. 
%and $\varepsilon^*=\beta \varepsilon$ is the dimensionless interaction strength, with $\beta$ the inverse thermal energy $\beta=(k_B T)^{-1}$.
Further, the dipolar interactions are described by the usual (three-dimensional) dipole-dipole potential,

\begin{equation}
\begin{split}
 u_{dd}(\textbf{r}_{ij},\boldsymbol{\mu}_i,\boldsymbol{\mu}_j)  &= \frac{\boldsymbol{\mu}_i \cdot \boldsymbol{\mu}_j}{r_{ij}^3} - 3\frac{\left( \boldsymbol{\mu}_i \cdot \textbf{r}_{ij} \right) \left( \boldsymbol{\mu}_j \cdot \textbf{r}_{ij} \right)}{r_{ij}^5}  
\end{split}
\label{brown-dyn-mod_eq:4}
\end{equation}
which is long-ranged ($\sim r^{-3}$) and non-separable, that is, in $ u_{dd}$ the spatial configuration of the dipole moments is coupled to their orientation. For the theoretical calculations presented in \cref{theoreticaldesc}, the long-range character does not impose a problem, since we are considering a two-dimensional system and thus, all spatial integrals converge. However, for the numerical (Brownian dynamics) simulations, the long-range character necessitates the use of special techniques (accompanied by larger computational cost) in order to avoid any truncation-induced bias affecting, e.g., correlation functions \cite{mazars2011}. Here we use the 2D Ewald summation technique \cite{Liao2020, Liao2020correction}. Finally, the non-separability requires some more care in the evaluation of angular integrals, as we will discuss in \cref{theoreticaldesc,2-bodycorrelfunct}. 

\section{Coarse-grained description}\label{theoreticaldesc}

\subsection{Derivation of the effective hydrodynamic equations}
 Following \cite{Sese-Sansa2021}, we here derive hydrodynamic equations based on a Fokker-Planck approach. We start from the $N$-body Smoluchowski equation, corresponding to the Langevin equations, \cref{brown-dyn-mod_eq:1,brown-dyn-mod_eq:2}, which account for the time evolution of the joint probability distribution $\psi_{N}(\{\textbf{r}_{i}, \varphi_{i}\};t)$,
% . Following \cite{Bialke2013, Sese-Sansa2021}, we rewrite the over-damped Langevin dynamics, \cref{brown-dyn-mod_eq:1,brown-dyn-mod_eq:2}, in the form of an $N$-body Smoluchowski equation,

\begin{equation}
\begin{split}
\partial_{t} \psi_{N}& = \sum_{i=1}^{N} \nabla_{i} \cdot \left(  \beta D_t (\nabla_{i} U) \psi _{N} - v_{0} \hat{\textbf{e}}_{i} \psi _{N} + D_t \nabla_{i} \psi _{N} \right)  \\
 & \qquad  +\sum_{i=1}^{N} \partial _{\varphi_{i}} \left( \beta D_r \left(\partial_{\varphi_{i}} U \right) \psi _{N}+D_{r} \partial_{\varphi_{i}} \psi_{N} \right).
\label{deriv_hydro-eq:1}
\end{split}
\end{equation}
Here, $U$ is the full interaction potential given in \cref{brown-dyn-mod_eq:3}. 
%Particles self-propel at constant speed $v_0$ and are subjected to translational and rotational diffusion.
%From the N-body SE, a standard approach to reduce the number of v to one can reduce the number of variables by integration over 
 %it is possible to integrate out $N-1$ variables, $\psi_{1} (\textbf{r}_{1},\varphi_{1};t) = N \int_{-\infty}^{\infty} d\textbf{r}_{2} ... d\textbf{r}_{N} \int_{0}^{2\pi} d\varphi_{2} ... d\varphi_{N}  \psi_{N}$, to obtain a Smoluchowski equation for the 1-body distribution,
One can then obtain the one-body distribution by integrating out $N$-1 positional and angular variables, $\psi_{1} (\textbf{r}_{1},\varphi_{1};t) = N \int_{-\infty}^{\infty} d\textbf{r}_{2} ... d\textbf{r}_{N} \int_{0}^{2\pi} d\varphi_{2} ... d\varphi_{N}  \psi_{N}$. Assuming indistinguishability of particles, this procedure yields the one-body Smoluchowski equation,

\begin{equation}
\begin{split}
\partial_{t} \psi_{1} &= - \nabla_{1} \cdot \left( \beta D_t  \textbf{F} \left(\textbf{r}_{1},\varphi_{1};t\right)+ v_{0} \hat{\textbf{e}}_1 \psi _{1} - D_t \nabla_{1} \psi _{1} \right) \\
&\qquad  - \frac{\partial}{\partial \varphi_{1}} \left(  \beta D_r T \left(\textbf{r}_{1},\varphi_{1};t \right)- D_{r} \frac{\partial \psi_{1}}{\partial \varphi_{1}} \right).
 \end{split}
\label{deriv_hydro-eq:2}
\end{equation}

In \cref{deriv_hydro-eq:2} we have introduced the effective force, $\textbf{F} (\textbf{r}_{1},\varphi_{1};t)$, and the scalar torque, $T (\textbf{r}_{1},\varphi_{1};t)$, that encode the pair-wise interactions between the tagged particle (labeled \textit{1}) and the surrounding particles. Specifically, the force is given by

%ALL THE DEPENDENCY ON TWO-PARTICLE CORRELATIONS IS IN 
%where particles are assumed to be indistinguishable, all the pairwise interactions are encoded in the force, $\textbf{F} (\textbf{r}_{1},\varphi_{1};t)$, and torque, $T (\textbf{r}_{1},\varphi_{1};t)$, which are the effective force and torque that the surrounding particles exert on the tagged particle (labeled \textit{1}). Their full expression reads,

\begin{equation}
\begin{split}
\textbf{F}& (\textbf{r}_{1},\varphi_{1};t)\\
 &=-N \int_{-\infty}^{\infty} d\textbf{r}_{2} ... d\textbf{r}_{N} \int_{0}^{2\pi} d\varphi_{2} ... d\varphi_{N} \left(\nabla_{1} U \right)\psi _{N} \\
&=-\int_{-\infty}^{\infty} d\textbf{r}_{2} \int_{0}^{2\pi} d\varphi_{2} \frac{\partial u_{sr} \left(r_{12}\right)}{\partial \textbf{r}_{1}} \psi _{2}\left(\textbf{r}_1,\textbf{r}_2,\varphi_1,\varphi_2; t \right) \\
&- \int_{-\infty}^{\infty} d\textbf{r}_{2} \int_{0}^{2\pi} d\varphi_{2} \frac{\partial u_{dd}(\textbf{r}_{12},\boldsymbol{\mu}_1,\boldsymbol{\mu}_2)}{\partial \textbf{r}_{1}}\psi _{2} \left(\textbf{r}_1,\textbf{r}_2,\varphi_1,\varphi_2; t \right), 
\end{split}
\label{deriv_hydro-eq:3}
\end{equation}
where we have used \cref{brown-dyn-mod_eq:3}, and $\psi _{2} \left(\textbf{r}_1,\textbf{r}_2,\varphi_1,\varphi_2; t \right) $ is the two-body probability density. While the force involves the contributions arising from steric and dipole-dipole interactions, the torque arises solely due to the dipolar coupling and reads

\begin{equation}
\begin{split}
&T (\textbf{r}_{1},\varphi_{1};t) \\
&= -N \int_{-\infty}^{\infty} d\textbf{r}_{2} ... d\textbf{r}_{N} \int_{0}^{2\pi} d\varphi_{2} ... d\varphi_{N} \frac{\partial U}{\partial \varphi_{1}}\psi _{N}\\
&=- \int_{-\infty}^{\infty} d\textbf{r}_{2} \int_{0}^{2\pi} d\varphi_{2}  \frac{\partial u_{dd}(\textbf{r}_{12},\boldsymbol{\mu}_1,\boldsymbol{\mu}_2)}{\partial \varphi_{1}}\psi _{2} \left(\textbf{r}_1,\textbf{r}_2,\varphi_1,\varphi_2; t \right).
\end{split}
\label{deriv_hydro-eq:4}
\end{equation}
%$\boldsymbol{\mu}_1 = \mu \hat{\boldsymbol{e}}_1=\mu(\cos \varphi_1, \sin \varphi_1)$
%Note that steric effects derive from a central force field, only depending on interparticle distances $r_{12} = |\boldsymbol{r}_2 - \boldsymbol{r}_1|$. However, dipolar interactions, $ u_{dd}(\textbf{r}_{12},\boldsymbol{\mu}_1,\boldsymbol{\mu}_2)$, couple positions and orientations, where we take the dipolar moment to be parallel to the direction of self-propulsion at all times, as stated in \cref{modelmethods}. 

To proceed, it is useful to define appropriate independent variables. A standard choice is to express all vectors $(\textbf{r}_{12},\boldsymbol{\mu}_1,\boldsymbol{\mu}_2)$ via their polar angles in a laboratory frame of reference,  such that the system is fully defined by the set of variables
$(r_{12},\omega,\varphi_1,\varphi_2)$ with $\mathbf{r}_{12}=\mathbf{r}_2-\mathbf{r}_1=r_{12}(\cos\omega,\sin\omega)$. Henceforth, we focus on situations where the system is homogeneous and globally isotropic, i.e., there is no global symmetry breaking. In such a situation, we can equivalently employ a body-fixed frame where all angles are expressed relative to the (arbitrary) direction of $\mathbf{r}_{12}$. To this end, we introduce the variables
$\theta_1=\varphi_1-\omega$, $\theta_2=\varphi_2-\omega$, and $\varphi_{12}=\varphi_2-\varphi_1$.
Note that since $\theta_2$ can be expressed via $\theta_1$ and $\varphi_{12}$, i.e., $\theta_2=\varphi_{12}-\theta_1$, we henceforth use $\theta_1$ and $\varphi_{12}$ as independent variables.

\begin{figure}
\centering
  \includegraphics[trim=0 0 0 0, width=0.6\columnwidth]{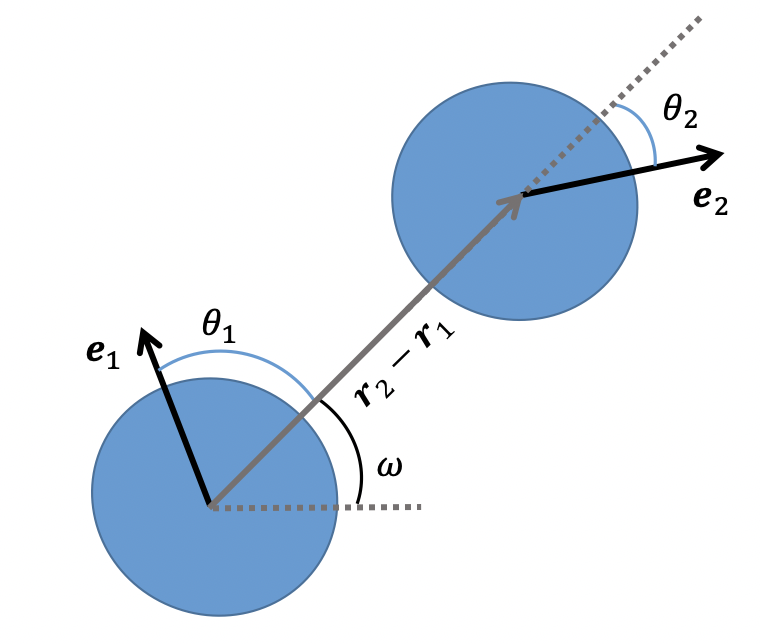}
  \caption{Sketch of the system's variables.}
  \label{brown-dyn-mod_fig:1}
\end{figure}

We now consider the integrals appearing in the expressions for the effective force and torque (\cref{deriv_hydro-eq:3,deriv_hydro-eq:4}).
Both expressions involve integrals over $\varphi_2$ and the direction of $\mathbf{r}_2$, while the coordinates of particle $1$, $\mathbf{r}_1$ and $\varphi_1$, are kept fixed.
Having this in mind, we can replace the integral over $\varphi_2$ by an integral over $\varphi_{12}$. Further, by setting $\mathbf{r}_1$ into the origin (which can be done due to the overall homogeneity of the state considered),
the integral over $\mathbf{r}_2$ can be expressed as 
$\int d\mathbf{r}_2\ldots=\int d\mathbf{r}_{12}\ldots=\int dr_{12}r_{12}\int d\omega\ldots=-\int dr_{12} r_{12} \int d\theta_1\ldots$,
where we have used that $\theta_1=\varphi_1-\omega$ and $\varphi_1$ is kept fixed.

The same angular variables can also be used to express the angular dependencies of correlation functions, see \cref{2-bodycorrelfunct}. Specifically, following \cite{Bialke2013}, we decompose the two-body probability density as
 
\begin{equation}
\psi_{2}(\textbf{r}_1,\textbf{r}_2,\varphi_{1},\varphi_{2};t) = \psi_{1}(\textbf{r}_1,\varphi_{1};t) \bar{\rho} G(r_{12},\theta_1,\varphi_{12};t),
\label{deriv_hydro-eq:8}
\end{equation}
where  $\bar{\rho}$ is the mean density and all the structural information contained in the two-body correlations has been cast into the correlation function $G(r_{12},\theta_1,\varphi_{12};t)$.
Inserting \cref{deriv_hydro-eq:8} into the equation for the force, and projecting on the orientation of particle $1$, we obtain 
%Introducing this decomposition into the equation for the force projected along $\hat{\textbf{e}}_1$ and regrouping the 1-body and 2-body terms leads to
 $\hat{\textbf{e}}_1 \cdot \textbf{F} (\textbf{r}_{1},\varphi_{1};t)= - \psi_{1}(\textbf{r}_1,\varphi_{1};t) \bar{\rho} \zeta$, where
%where $\psi_{1}$ is the 1-body probability distribution function, is the average density and $G(r_{12},\varphi_{12},\theta_1,t)$ is the pair-correlation function encoding the microscopic properties of the system. 

%We interpret it as the probability to find a particle (labeled \textit{ 2}) with orientation $\varphi_2$ and at a distance $r_{12}$ from the tagged particle (labeled \textit{1}), with orientation $\varphi_1$ and relative in-plane orientation $\theta_1$, see \cref{brown-dyn-mod_fig:1}.

%We now proceed by projecting the force $\textbf{F} (\textbf{r}_{1},\varphi_{1};t)$ into $\hat{\textbf{e}}_1$, to obtain the force component acting in the direction of self-propulsion of the tagged particle. We group all the dependence on pair-wise interactions in an effective coefficient, $\zeta$,

%\begin{equation}
%\begin{split} 
%\hat{\textbf{e}}_1 \cdot \textbf{F} = - \psi_{1}(\textbf{r}_1,\varphi_{1},t) \bar{\rho} \zeta
%\end{split}
%\label{deriv_hydro-eq:11}
%\end{equation}

\begin{equation}
\begin{split} 
\zeta&=\int_{0}^{\infty}  d r_{12} r_{12}  \int_{0}^{2\pi}  d\theta_1 \cos \theta_1\\
&\qquad \int_{0}^{2\pi} d\varphi_{12}  \left(- \frac{\partial u_{sr} \left(r_{12}\right)}{\partial r_{12}} \right) G(r_{12},\theta_1,\varphi_{12};t) \\
&+ \int_{0}^{\infty}  d r_{12} r_{12}  \int_{0}^{2\pi}  d\theta_1 \\
&\qquad \int_{0}^{2\pi} d\varphi_{12} 3 \mu^2  \Big( \frac{\cos \left(\varphi_{12} + \theta_1\right)}{r_{12}^4} +\frac{\cos \theta_1}{r_{12}^4} \cos \varphi_{12} \\
&+ \frac{\cos \varphi_{12} - 5  \cos \theta_1 \cos \left(\varphi_{12} + \theta_1\right) }{r_{12}^4} \cos \theta_1 \Big)G(r_{12},\theta_1,\varphi_{12};t).
\end{split}
\label{deriv_hydro-eq:12}
\end{equation}
We note that the introduction of the parameter $\zeta$ is in accordance with earlier studies \cite{Bialke2013,Sese-Sansa2021}. The idea is to express the vectorial force $\textbf{F}$ appearing in \cref{deriv_hydro-eq:3} in terms of an approximate basis spanned by $\hat{\textbf{e}}_1$ and $\nabla \psi_1$, $\textbf{F}  \approx (\hat{\textbf{e}}_1 \cdot \textbf{F})\hat{\textbf{e}}_1 + \left(\frac{\left(\nabla_1 \psi_1-(\hat{\textbf{e}}_1 \cdot \textbf{F})\hat{\textbf{e}}_1 \right)\cdot  \textbf{F}}{|\nabla_1 \psi_1|^2}\right)\nabla_1 \psi_1$ (see \cite{Sese-Sansa2021} for details). The quantity $\zeta$ can then be interpreted as a translational friction in the direction of self-propulsion.

In a similar manner, the torque defined in \cref{deriv_hydro-eq:4} can be rewritten as $T (\textbf{r}_{1},\varphi_{1};t) = -  \psi_{1}(\textbf{r}_1,\varphi_{1};t) \bar{\rho} \varepsilon$, where

%encoding all the pair-wise interactions in an effective coefficient, $\varepsilon$,

%\begin{equation}
%\begin{split} 
%T = -  \psi_{1}(\textbf{r}_1,\varphi_{1},t) \bar{\rho} \varepsilon
%\end{split}
%\label{deriv_hydro-eq:13}
%\end{equation}

\begin{equation}
\begin{split} 
 \varepsilon =  \int_{0}^{\infty}  &d r_{12} r_{12}  \int_{0}^{2\pi}  d\theta_1 \int_{0}^{2\pi} d\varphi_{12} \mu^2 \left(  \frac{\sin ( \varphi_{12})}{r_{12}^3} \right. \\
 &\left. + 3 \frac{ \sin  \theta_1  \cos \left(\varphi_{12} + \theta_1\right)}{r_{12}^3} \right) G(r_{12},\theta_1,\varphi_{12};t).
\end{split}
\label{deriv_hydro-eq:14}
\end{equation}
In analogy to $\zeta$, $\varepsilon$ can be interpreted as a rotational friction coefficient \cite{Sese-Sansa2021}. Inserting the above expressions into the 1-body Smoluchowski equation we obtain

\begin{equation}
\begin{split}
\partial_{t} \psi_{1} &= - \nabla_{1} \cdot \left(  \left(v_{0} -\beta D_t \bar{\rho} \zeta \right) \hat{\textbf{e}}_1 \psi _{1} - \Da \nabla_{1} \psi _{1} \right) \\
&\qquad  + \frac{\partial}{\partial \varphi_{1}} \left(  \beta D_r \bar{\rho} \varepsilon  \psi_1  + D_{r} \frac{\partial \psi_{1}}{\partial \varphi_{1}} \right),
 \end{split}
\label{deriv_hydro-eq:16}
\end{equation}
%= (\hat{\textbf{e}}_1 \cdot \textbf{F})\hat{\textbf{e}}_1 + \delta \textbf{F}
where we have defined an effective translational diffusion coefficient  $\Da = D_t - \beta D_t \frac{\left(\nabla_1 \psi_1-(\hat{\textbf{e}}_1 \cdot \textbf{F})\hat{\textbf{e}}_1 \right)\cdot  \textbf{F}}{|\nabla_1 \psi_1|^2}$, which includes a term deriving from Gram-Schmidt orthonormalization (see \cite{Sese-Sansa2021} for further details). 

Following the approximation first introduced in \cite{Bialke2013}, we consider  $\Da$ to be a constant corresponding to the long-time diffusion coefficient of a suspension of passive particles. Consequently, all the dependency of \cref{deriv_hydro-eq:16} on 2-body correlations is now encoded in the effective coefficients $\zeta$ and $\varepsilon$, which appear in the advective term of the translational and rotational degrees of freedom, respectively (i. e., the 1st and 3rd term on the right hand side of \cref{deriv_hydro-eq:16}). 

%We now perform a Gram-Schmidt orthonormalization on the force $\textbf{F}$ to decompose it in the vector basis formed by the direction of self-propulsion and the gradient of the probability distribution, $\{ \hat{\textbf{e}}_1, \nabla_1 \psi_1\}$. 

%\begin{equation}
%\begin{split} 
%\textbf{F}  \approx (\hat{\textbf{e}}_1 \cdot \textbf{F})\hat{\textbf{e}}_1 + \left(\frac{\left(\nabla_1 \psi_1-(\hat{\textbf{e}}_1 \cdot \textbf{F})\hat{\textbf{e}}_1 \right)\cdot  \textbf{F}}{|\nabla_1 \psi_1|^2}\right)\nabla_1 \psi_1
%\end{split}
%\label{deriv_hydro-eq:15}
%\end{equation}

\cref{deriv_hydro-eq:16}  may be considered as the first member of the BBGKY-hierarchy, relating the one-body distribution to 2-body correlations via the effective coefficients  $\zeta$ and $\varepsilon$.
We now consider the effective coefficients to be spatially constant parameters, $\zeta = \zeta_0$ and $\varepsilon= \varepsilon_0$, which is plausible in the overall homogeneous, isotropic phase. For the subsequent theoretical analysis we moreover assume $\zeta_0$ and $\varepsilon_0$ to be independent of the actual strength and shape of correlations. This implies an additional (mean-field like) approximation, which closes the hierarchy of coupled equations \cite{Bialke2013, Sese-Sansa2021}. We stress, however, that a direct mapping to simulations remains possible by numerically computing $\zeta$ and $\varepsilon$ from the Langevin dynamics, which we will later do to test the theoretical prediction (see \cref{mapping_meanfield_micro_model}). To ease notation, from now on, we drop the subscripts in $\zeta_0$ and $\varepsilon_0$ to denote the constant mean-field coefficients. 
%To close the hierarchy, we perform a mean-field approximation by considering such coefficients to be constant parameters, $\zeta = \zeta_0$ and $\varepsilon= \varepsilon_0$, therefore neglecting their dependence on pairwise correlations \cite{Bialke2013, Sese-Sansa2021}. 

To proceed in our coarse-grained theory, we consider the lowest-order moments of $\psi_1$ with respect to $\hat{\textbf{e}}_1$. In the remainder of the paper, we drop the numerical subscripts designating particles. The zeroth moment corresponds to the density field, $ \rho (\textbf{r},t) \equiv \int_{0}^{2 \pi} d \varphi \psi_{1} (\textbf{r},\varphi,t) $, and the first moment to the polarization, $  \textbf{p} (\textbf{r},t) \equiv \int_{0}^{2 \pi} d \varphi \hat{\textbf{e}} \psi_{1} (\textbf{r},\varphi,t)$.
Integrating \cref{deriv_hydro-eq:16}, we thereby obtain the effective hydrodynamic equations for each of the two fields,

\begin{equation}
\begin{split}
 \partial_{t} \rho (\textbf{r},t) =- \nabla \cdot \Big( v \textbf{p} -\Da \nabla \rho \Big),
\end{split}
\label{deriv_hydro-eq:20}
\end{equation}

\begin{equation}
\begin{split}
 \partial_{t} \textbf{p} (\textbf{r},t) = -\nabla& \cdot \Big(v (\frac{1}{2}  \rho  \mathbbm{1}+ \textbf{Q} )- \Da \nabla \textbf{p} \Big) \\
 &- \beta D_r \bar{\rho} \varepsilon_0 \textbf{p}^{\perp} - D_{r} \textbf{p},
 \end{split}
\label{deriv_hydro-eq:21}
\end{equation}
where the perpendicular vector $\mathbf{p}^{\perp}$ follows as $\mathbf{p}^{\perp}=\mathcal{R} \mathbf{p}$, with  $\mathcal{R} = \left(\begin{array}{ccc} 0 & -1\\1 & 0\\ \end{array}\right)$. Note that each hydrodynamic equation is coupled to its next order moment. In particular, $\textbf{p}$ is coupled to the tensor $ \textbf{Q} $. Here, we set $ \textbf{Q} = 0$ \cite{Cates_runtumble2013,VanDamme2019,Zhang2021,Bialke2013,Sese-Sansa2021}
and thereby obtain a closed set of hydrodynamic equations describing the evolution of the density field and the polarization. As shown in \cite{Sese-Sansa2021} and also discussed below, this is a good approximation to study the linear destabilitzation leading to MIPS within the globally isotropic state. 

The hydrodynamic equations (14) and (15) %\cref{deriv_hydro-eq:20,deriv_hydro-eq:21} 
describe systems of self-propelled disks subject to certain conservative forces and torques. In the present model, torques derive from dipole-dipole interactions. It should be noted, however, that the functional form of the equations is the same than for systems of ABP interacting via simpler velocity-alignment rules \cite{Sese-Sansa2021}. This highlights the generality of our approach, where the microscopic structure enters only via the effective friction coefficients $\zeta$ and $\varepsilon$.

\subsection{Linear stability analysis} 

We now study the onset of motility-induced phase separation of a homogeneous, globally isotropic (i. e., $\textbf{p} = 0$) suspension of dipolar active particles by means of a linear stability analysis. To this end, we consider tiny perturbations to the homogeneous and isotropic solution of the hydrodynamic equations, $\rho (\textbf{r}) = \bar{\rho} + \delta \rho$, $\textbf{p}(\textbf{r}) =  \delta \textbf{p}$, and calculate their evolution via \cref{deriv_hydro-eq:20,deriv_hydro-eq:21}  up to linear order in $\delta \rho$ and $\delta \textbf{p}$.

Due to the generality of our coarse-grained model, the predictions of the linear stability analysis for a system of dipolar ABP will be formally identical to those for systems of ABP with Vicsek-like aligning rules. It is for this reason that we refer to \cite{Sese-Sansa2021} for a detailed description of the linear stability analysis. Here, we just mention the most relevant steps.
 
At the level of our hydrodynamic description, motility-induced phase separation, which is a \textit{macroscopic} phase separation, is identified as a destabilization of the homogeneous and isotropic phase, in the limit $\textbf{q} \rightarrow 0$, where $\textbf{q}$ is the wave vector of the perturbation. In other words, MIPS corresponds to a long wavelength instability \cite{Bialke2013,Sese-Sansa2021}.
As shown in \cite{Sese-Sansa2021}, $\rho(\textbf{r},t)$ is the slowest moment of the probability distribution $\psi_1$ and higher order moments are enslaved to $\rho(\textbf{r},t)$. This feature leads, without loss of generality, to the possibility of rewriting the evolution equations for the perturbation, $(\delta \rho, \delta \textbf{p})$, as $\partial_{t} \delta \hat{ \rho}(\textbf{q})  =  \Da^{eff}_{\textbf{q}} \textbf{q}^2 \delta \hat{ \rho}(\textbf{q}) $, where $\hat{*}$ denotes the Fourier transform.
In the long wavelength limit, the effective diffusion coefficient reads $ \Da^{eff}_{0}= \frac{1}{2} \left(v_0 - \bar{\rho} \zeta) (v_0 -2 \bar{\rho}  \zeta \right) \frac{D_r}{D_r^2 + (\bar{\rho} \varepsilon)^2} - \Da  $. A spinodal-line instability of the homogeneous and isotropic phase corresponds to $\Da^{eff}_{0} < 0$. Therefore, the stability limit is given by $\Da^{eff}_{0} = 0$, which leads to the boundaries

\begin{equation}
\begin{split}
\zeta^{\pm}= \frac{3 v_0}{4 \bar{\rho}} \pm \frac{1}{4\bar{\rho}}\sqrt{v_0^2-16\Da D_r -16\Da \frac{\left(\bar{\rho} \varepsilon \right)^2}{D_r} }.
\end{split}
\label{modeldescrip_eq:22}
\end{equation}

To investigate the occurrence of MIPS, we thus have to calculate the limit of the instability region, $\zeta^{-} <\zeta < \zeta^{+}$, as a function of the system's activity, $v_0$. \cref{modeldescrip_eq:22} can be written in dimensionless units by defining the reduced self-propulsion speed $\frac{v_0}{v^*}$, where $v^*=4\sqrt{\Da D_r}$. The translational and rotational friction coefficients in their dimensionless form are $\tilde{\zeta} =\frac{\bar{\rho}}{v^*}\zeta$ and $\tilde{\varepsilon} =\frac{\bar{\rho}}{D_r}\varepsilon$. We will employ these dimensionless quantities in the subsequent section, when comparing the prediction of the mean-field coarse-grained model with those from particle-resolved simulations.

In the remainder of the derivation, we set $\tilde{\varepsilon}=0$. We stress that this is not an approximation, but can be argued by symmetry arguments based on the dipolar pair interactions considered here and on the fact that we are considering a globally isotropic phase. We refer to \cref{appxepscoeff} for a detailed explanation and numerical confirmation.

%Here, we have assumed the local density $\rho(\textbf{r},t)$ to be a slow varying field, which is reasonable in the homogeneous state. Therefore, the effective self-propulsion velocity is now $v(\rho) = v_0 -\beta D_t \rho \zeta_0 $. 
\section{Numerical results}\label{numeric_results}
%In \cref{theoreticaldesc} we have obtained a theoretical prediction for the limit of stability of the homogeneous and isotropic phase, \cref{modeldescrip_eq:22}. We would like now to test this prediction agains simulations. With the coarse-grained description derived above, this is a straightforward task, since the dependence of the theory with the microscopic model is captured in the effective coefficient $\zeta$, \cref{deriv_hydro-eq:12}. It is for this reason that we have performed Brownian dynamics simulations of the model system described in \cref{modelmethods}, \cref{brown-dyn-mod_eq:5,brown-dyn-mod_eq:4}, which will allow us to compute numerically $\zeta$. See \cref{browniandyndetails} for the simulation details. 

%Here, we recall that the rotational friction coefficient is $\varepsilon=0$, as explained in \cref{appxepscoeff}. Therefore, $\zeta$ is the parameter that captures all the relevant information concerning phase separation, evidencing that the aggregation mechanism that eventually leads to phase separation (in the weak coupling regime) is controlled by the competition between swimming persistence and excluded-volume interactions.

In \cref{theoreticaldesc} we have obtained a theoretical prediction for the limit of stability of the homogeneous and isotropic phase in terms of boundary values for the coefficient $\tilde{\zeta}$, see \cref{modeldescrip_eq:22}. The envisioned instability has the character of unstable (long wavelength) density fluctuations, that is, we are focusing on the onset of MIPS rather than on, e.g., a flocking state characterized by long-range polarization.
We stress here (again) that the stability limit has been derived assuming that $\tilde{\zeta}$ does not depend on details of the microscopic interaction; in fact, we have put $\tilde{\zeta}$ to a constant.
Thus, the stability limit only assumes that the two-body correlation function is anisotropic (due to activity), otherwise $\tilde{\zeta}$ would vanish from scratch. We recall, however, that our full hydrodynamic theory is linked to the type of microscopic interaction and resulting structure via \cref{deriv_hydro-eq:12}.

The goal is now to evaluate the performance of the theoretical prediction \cref{modeldescrip_eq:22} for a system of dipolar ABP.
To this end, we proceed as follows. We first discuss in \cref{browndyn_simus} results from direct numerical simulations of the Langevin \cref{brown-dyn-mod_eq:1,brown-dyn-mod_eq:2}, focusing on the identification of parameters where MIPS occurs without breaking the rotational symmetry. Technical details of these simulations are given in \cref{browniandyndetails}.
Secondly, we investigate properties of the numerically obtained correlation functions and use these to provide results for the coefficient $\tilde{\zeta}$ (\cref{mapping_meanfield_micro_model}). By investigating the dependency of $\tilde{\zeta}$ on activity (i.e., on $v_0^{*}$), we can eventually compare with the theoretical prediction, \cref{modeldescrip_eq:22}.
We recall that investigating $\tilde{\zeta}$ is indeed sufficient since the "rotational friction" coefficient $\tilde{\varepsilon}$ vanishes by symmetry in the globally isotropic phase (see \cref{appxepscoeff}).

\subsection{Brownian Dynamics simulations}\label{browndyn_simus}
%Increasing the dipolar coupling strength, the system develops global orientational order, as already observed in \cite{Liao2020}. The structures in the ordered state are motile and thus their formation cannot be attributed to the self-trapping mechanism leading to MIPS. This is the reason why it is important to first identify the regime of weak dipolar coupling, in which the system does not develop a global net polarization. 

Our first goal is to identify parameters where MIPS occurs within the globally isotropic phase ($\textbf{p}=0$). Clearly, crucial parameters are the packing fraction $\phi$ (which we here define as $\phi=N\pi R^2/(4L^2)$), the P\'eclet number (or motility)

\begin{equation}
v_0^* =v_0 \frac{ R}{D_t},
\label{eq_peclet}
\end{equation}
and the dipolar coupling strength

\begin{equation}
\lambda=\frac{\beta\mu^2}{R^3}.
 \label{eq_lambda}
\end{equation}
However, as it turns out, also the strength of the WCA repulsion, $u_0$, plays an important role. In an earlier simulation study of dipolar ABP \cite{Liao2020}, we have detected MIPS at a relatively high packing fraction of $\phi=0.63$, with $u_0=10k_B T$, in the range $\lambda\lesssim 0.71$. For the present paper, we have performed simulations at the somewhat lower packing fraction $\phi=0.4$ (and $u_0=100k_B T$). The reason for choosing a lower packing fraction is that one would generally expect the theoretical analysis to be the better, the smaller the density, and this is also confirmed by numerical results for pure ABP (see \cref{limits_continuum_theory}).
Thus, all results discussed subsequently pertain to $\phi=0.4$. The simulation parameters employed are detailed in \cref{browniandyndetails}.

\begin{figure}
\centering 
   	\includegraphics[trim=0 0 0 10, width=\columnwidth]{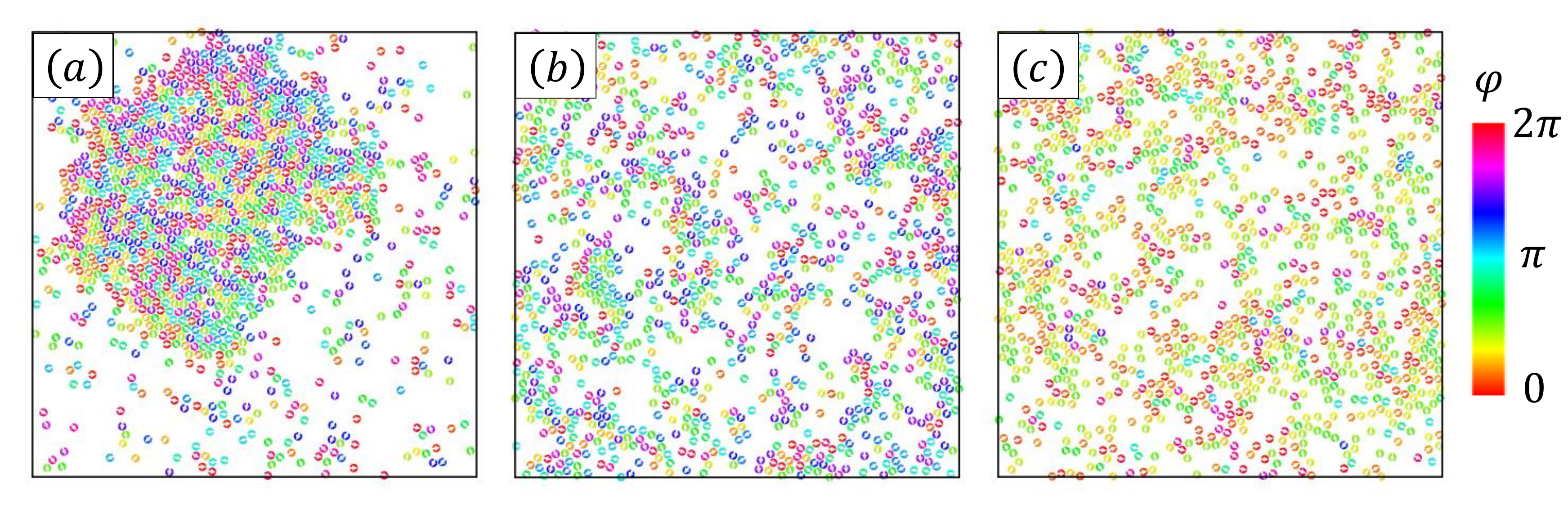}
\caption{Snapshots of the dipolar ABP system at $\phi=0.4$ and different combinations $(\lambda, v_0^*)$. (a) (0, 112.2); (b) (0.71, 112.2); (c) (1.41,112.2). The color code indicates the direction of self-propulsion which equals the direction of the permanent dipole moment.}
\label{results_fig:1}
\end{figure}

To give a first impression of the system's phase behavior as a function of $v_0^*$ and $\lambda$, we provide in \cref{results_fig:1} typical simulations snapshots of the system at different parameter regimes.
One clearly observes different morphologies, including clustered states, \cref{results_fig:1} (a), and flocking behavior, \cref{results_fig:1} (c), which will be discussed in detail below.

To start with, we determine the regime of "low" dipolar coupling where the system is globally isotropic. Indeed, as shown in \cite{Liao2020} and \cref{results_fig:1} (c), dipolar ABP can develop a net polarization (flocking) when $\lambda$ becomes sufficiently large.
To this end we compute the (scalar) global polarization,

\begin{equation}
\phi_e = \left\langle \left|\frac{1}{N}  \sum_{i=1}^N \hat{\textbf{e}}_i \right| \right\rangle, 
\label{results_eq:1}
\end{equation}
where $\langle ... \rangle$ denotes an ensemble average. An isotropic state with randomly distributed orientations is indicated by $\phi_e  \approx  0$, while large values of $\phi_e$ signal the emergence of an ordered polar state.

\begin{figure}
\centering
  \includegraphics[trim=0 0 0 0,width=\columnwidth]{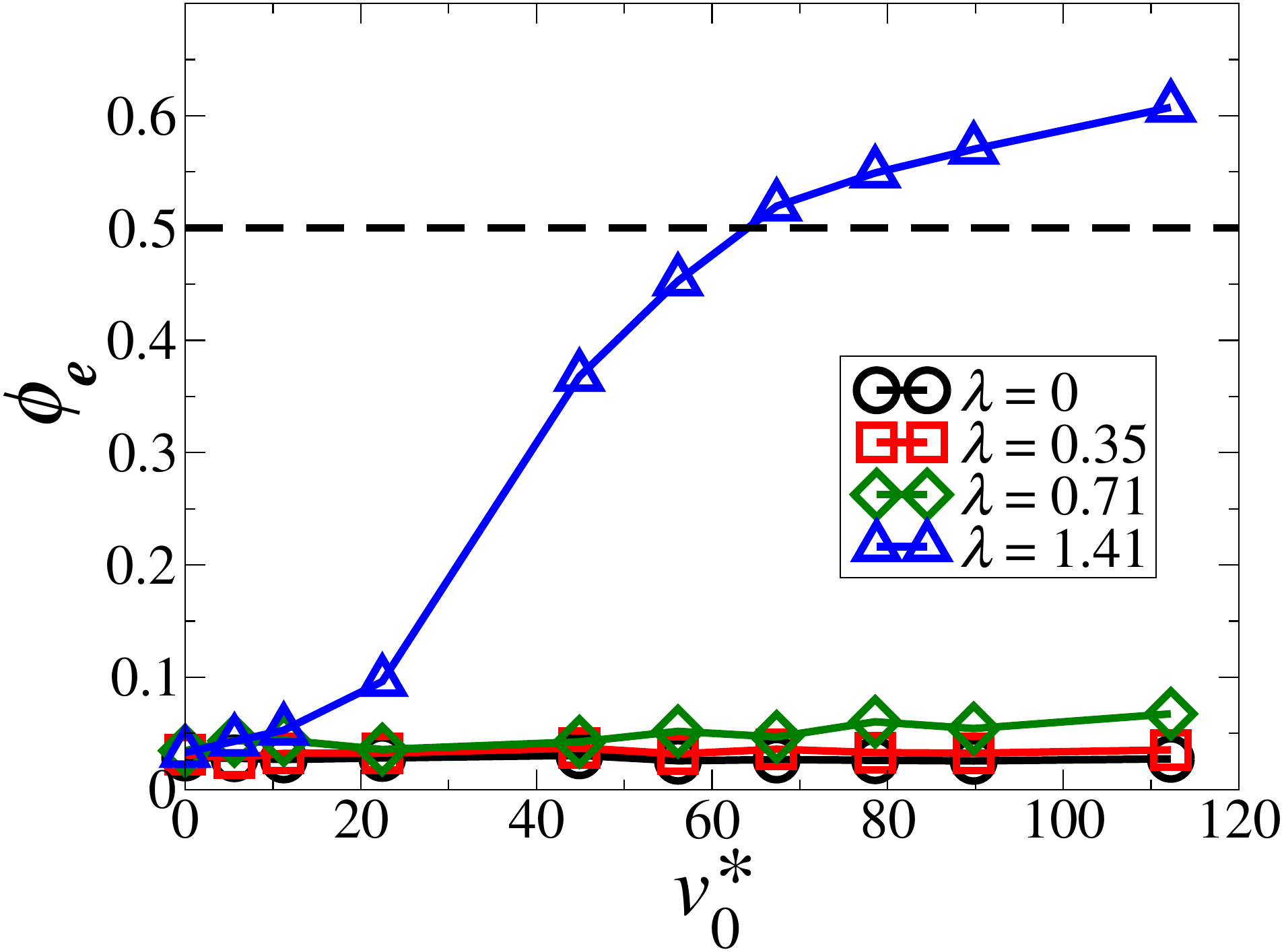}
  \caption{Global polarization as a function of the motility for different values of the dipolar coupling strength $\lambda$. }
   \label{results_fig:2.0} 
\end{figure}

In \cref{results_fig:2.0} we plot the global polarization as a function of the motility for different values of the dipolar coupling strength, $\lambda$. The data reveal that at $\lambda = 1.41$, the system transitions from a disordered state at low motilities to an orientationally ordered state at high values of $v_0^*$ (for an illustration of the corresponding microstructure see the snapshot in \cref{results_fig:1} (c)). On the contrary, below $\lambda = 0.71$, the polarization remains close to zero regardless of the self-propulsion speed, indicating that the system does not develop global orientational order (see also \cref{results_fig:1} (b)). We will thus focus on the regime $\lambda < 0.71$ throughout the rest of the paper. 

In what follows, we aim at addressing the effect of dipolar interactions on the onset of motility-induced phase separation. 
%We follow a two-fold approach. On the one hand, we analyse particle-based simulations to identify the onset of phase separation at different values of the dipolar coupling. On the other hand, we test the predictions of the coarse-grained model against simulations, \cref{mapping_meanfield_micro_model}. In order to do so, we take advantage of the fact that the hydrodynamic equations are linked to the microscopic model via the effective friction coefficient $\zeta$, which depends on the 2-body correlation function $G(r,\theta,\varphi)$, \cref{deriv_hydro-eq:12}. 
As a first indicator, we compute the probability for a particle to belong to the largest cluster of the system. We define a cluster as a set of particles whose center-to-center distance is closer than a certain threshold. Here, we fix this threshold to be the cutoff distance of the WCA potential, $R$. We note that the typical distance between two dipolar ABP (as measured from the radial distribution function) is always smaller than $R$ for all $\lambda$ considered, for details see \cref{appx_rad_dist_funct}. Therefore, our cluster criterion is not affected by $\lambda$. Based on these considerations, we count the number of particles in the largest cluster, $n_{lcl}$, and compute their average fraction as,

\begin{figure*}[ht]
 \centering
 \includegraphics[trim=550 0 550 0,width=\columnwidth]{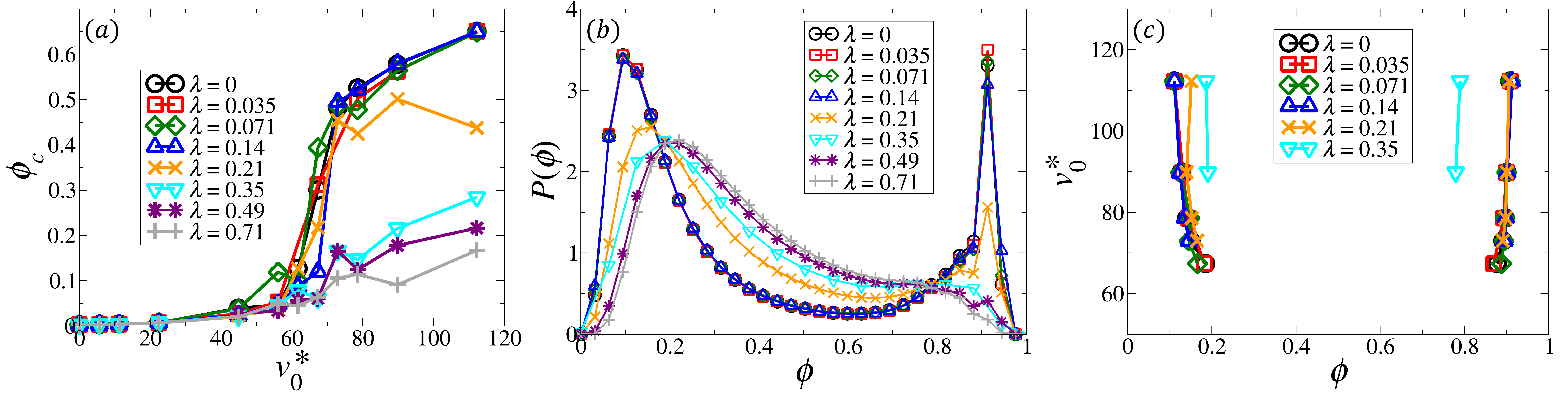}
 \caption{(a) Probability to belong to the largest cluster, $\phi_c$, as a function of the motility for different values of $\lambda$ pertaining to the isotropic state. (b) Probability distribution $P(\phi)$ of local area fractions at fixed motility $v_0^* = 112$ and varying $\lambda$. (c) Phase coexistence region for different values of $\lambda$.}
 \label{results_fig:2.1}
\end{figure*}

\begin{equation}
\phi_c = \frac{\langle n_{lcl} \rangle}{N}.
\label{results_eq:2}
\end{equation}

In \cref{results_fig:2.1} (a) we present results for $\phi_c $ as a function of the self-propulsion speed, $v_0^*$, for different values of (weak) dipolar coupling $\lambda$. One observes that the probability to belong to the largest cluster becomes finite upon increasing $v_0^*$, signaling a transition to a phase-separated state, see \cref{results_fig:1} (a) for illustration. This behavior is very familiar from other systems undergoing MIPS (see \cite{Fily2012,stenhammar2013,stenhammar2014,Cates2015,Solon2018,partridge2019}). Increasing $v_0^*$ leads to a mutual blocking of particles upon collision, triggering a feed-back mechanism by which more and more particles accumulate and further slow-down. 
%We point out that dipolar interactions increase the characteristic distance between particles in the regime of weak dipolar interactions, as seen from the radial distribution function in \cref{appx_rad_dist_funct}. However, such characteristic distance is always smaller than the threshold distance imposed to define a cluster. Therefore, we confirm that clusters are correctly identified and that the increase in characteristic distance induced by dipolar interactions does not interfere in the identification. 
From \cref{results_fig:2.1} (a) we observe that the largest cluster formed at high $v_0^*$ becomes smaller upon increasing $\lambda$. This phenomenon is well illustrated by \cref{results_fig:1} (b). Upon introducing a dipolar coupling strength $\lambda=0.71$ (which is close to the limit between the weak and strong coupling regimes), the MIPS cluster visible at $\lambda=0$ (\cref{results_fig:1} (a)) disappears and the system develops an isotropic, overall homogeneous phase, with local regions of higher density. This difference with respect to \cref{results_fig:1} (a) clearly indicates a suppression of MIPS due to dipolar interactions.  %This already indicates that dipole-dipole interactions tend to suppress the formation of large clusters, and thus, MIPS.

For a more quantitative evaluation of the impact of dipolar interactions on the motility-induced phase separation, we compute the regions of coexistence of a dense and a dilute phase, extracted from the probability distributions $P(\phi)$  of local area fractions (for more details of this method see \cite{Liao2020}).
\cref{results_fig:2.1} (b) shows these probability distributions at fixed motility $v_0^* = 112$ and varying dipolar interaction strength. At $\lambda = 0$, the system consists of one dense MIPS cluster coexisting with a dilute phase. This is reflected by a bimodal distribution with two sharp peaks. Specifically, the high density peak at $\lambda=0$ coincides with the hexagonal packing fraction, $\phi \approx 0.91$. Upon increasing $\lambda$, the low-density peak shifts to higher values of the local area fraction, indicating that the dilute phase becomes 'denser'. Moreover, the height of the high-density peak decreases until it transforms into the tail of a unimodal distribution. This occurs at $\lambda \approx 0.35$. For larger $\lambda$, the shape of $P(\phi)$ reveals that, even if there are regions where the density exceeds the mean average density, there is no full phase separation leading to the coexistence of a macroscopic cluster with a gas-like phase. Based on the probability distributions shown in \cref{results_fig:2.1} (b), we can extract the binodal curves related to MIPS by plotting the area fractions of the high- and low-density peaks in the $(\phi,v_0^*)$ plane, for different values of $\lambda$. This is done in \cref{results_fig:2.1} (c). We observe that upon increase of $\lambda$ the appearance of phase coexistence shifts towards higher values of $v_0^*$. Moreover, both the low- (high-) density branches move to higher (lower) density values of the packing fraction; that is, the coexistence region shrinks. Taken together, we see that the dipolar coupling indeed hinders the phase separation.
%This is in line with the behavior observed in $\phi_{c}(v_0^*)$, \cref{results_fig:2.1}.

\begin{figure}[ht]
\centering
  \includegraphics[width=\columnwidth]{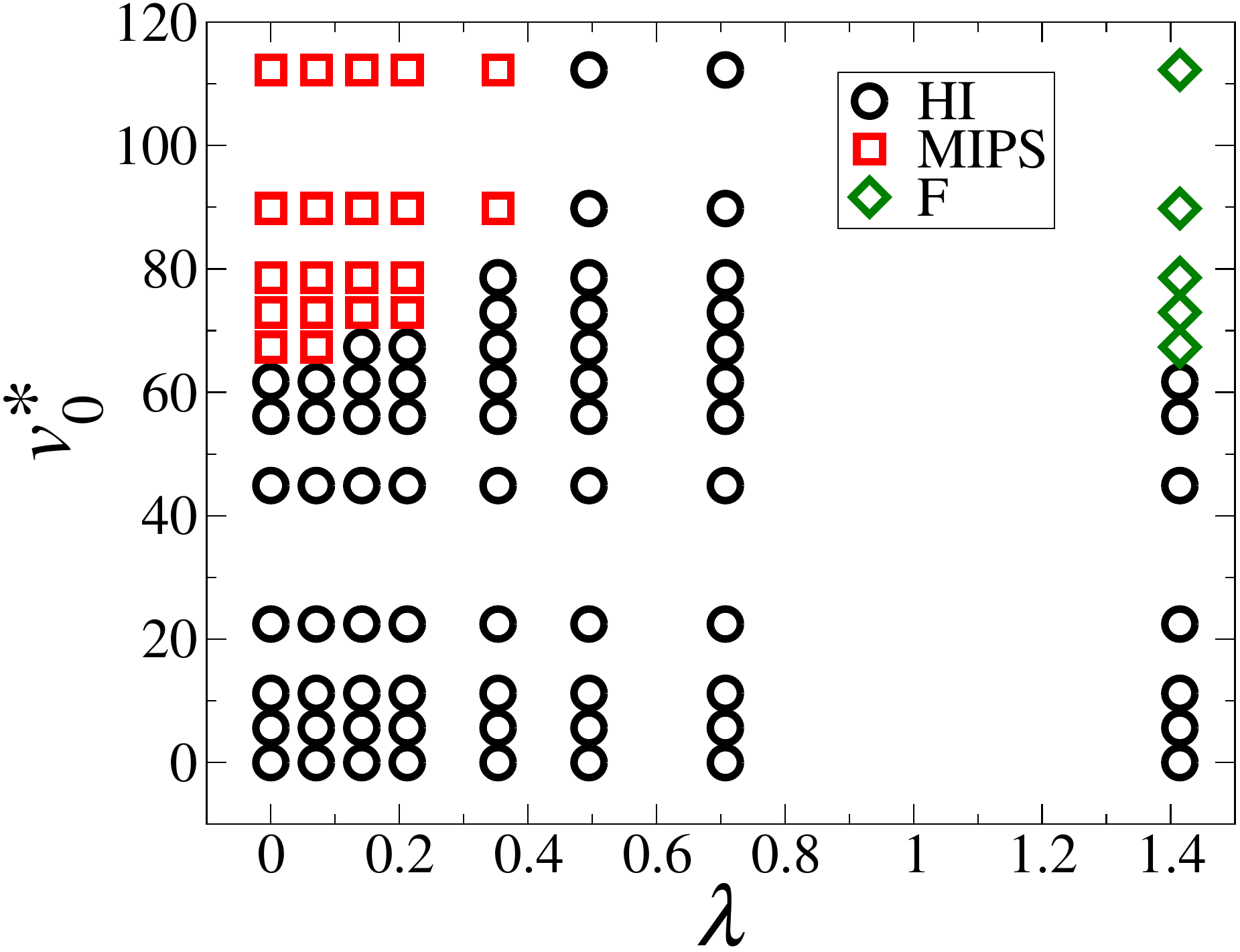}
  \caption{State diagram of a system of dipolar ABP in the $(\lambda, v_0^*)$ plane at fixed $\phi = 0.4$. Black circles indicate homogeneous states (HI), red squares correspond to MIPS states and green diamonds indicate flocking states (F), where a finite fraction of particles is oriented in the same direction ($\phi_e > 0$).}
   \label{results_fig:1.1} 
\end{figure}

To complete the picture, we provide in \cref{results_fig:1.1} a state diagram in the $(v_0^*,\lambda)$ plane, gathering all the information we have discussed so far. As outlined in \cref{browniandyndetails}, numerical simulations of dipolar active Brownian particles require substantial computational effort to take full account of the long-range nature of dipole-dipole interactions. It is for this reason that we have not explored the full state diagram, but have concentrated on the globally isotropic regime of weak coupling $(\lambda \le 0.71)$, which is the main region of interest in this work. Specifically, to locate the onset of MIPS at a given $\lambda$, we have searched for the motility $v_0^*$ where $P(\phi)$ changes from a unimodal to a bimodal shape. \cref{results_fig:1.1} then clearly reveals a shift of the phase separation to higher values of $v_0^*$ with increasing $\lambda$. This finding is consistent with earlier results for dipolar ABP at larger packing fraction \cite{Liao2020}. 
Closer inspection of the data presented in \cref{results_fig:1.1} shows that the shift is rather gradual in the range $\lambda\lesssim 0.2$. For larger $\lambda$, we observe an abrupt increase of the motility where MIPS sets in (see data point for $\lambda=0.35$). In other words, in this range, the system's behavior becomes very sensitive with respect to the strength of dipolar coupling, which may also explain the increasing numerical difficulties to reach convergent results. We will later see in \cref{mapping_meanfield_micro_model} that a similar abrupt change (at $\lambda\approx 0.35$) also occurs in the behaviour of the (numerically obtained) translational friction coefficient (\cref{mapping_theories_fig:1}).
Further, for $\lambda \geq 0.49$, there is no phase separation observed within the parameter range captured in \cref{results_fig:1.1}. Presumably, one has to go to even higher values of $v_0^*$ to see MIPS, but this was not systematically investigated in this study. We also identify a flocking state at high values of $\lambda$ and $v_0^*$, characterised by a finite polarization of $\phi_e > 0.5$, as obtained from \cref{results_fig:2.0}.

\subsection{Mapping of the mean-field theory onto the microscopic model}\label{mapping_meanfield_micro_model}
All in all, the results obtained from the particle-based (Brownian dynamics) simulations (see \cref{results_fig:1.1}) reveal that dipolar interactions hinder MIPS, as it was first reported in \cite{Liao2020} for a somewhat larger packing fraction. We now aim at connecting our particle-resolved results to the hydrodynamic theory described in \cref{theoreticaldesc}, particularly the prediction for the MIPS instability region, \cref{modeldescrip_eq:22}. 
As we have outlined in \cref{theoreticaldesc}, the link between the coarse-grained and the microscopic description is provided by the effective friction coefficient $\tilde{\zeta}$ that depends on the correlation function $G(r,\theta, \varphi,t)$, see \cref{deriv_hydro-eq:12}. From now on we focus on steady states, that is, $t\rightarrow\infty$, and thus drop the time argument in $G$.  

\begin{figure}
\centering
 \includegraphics[trim=0 0 0 5, width=\columnwidth]{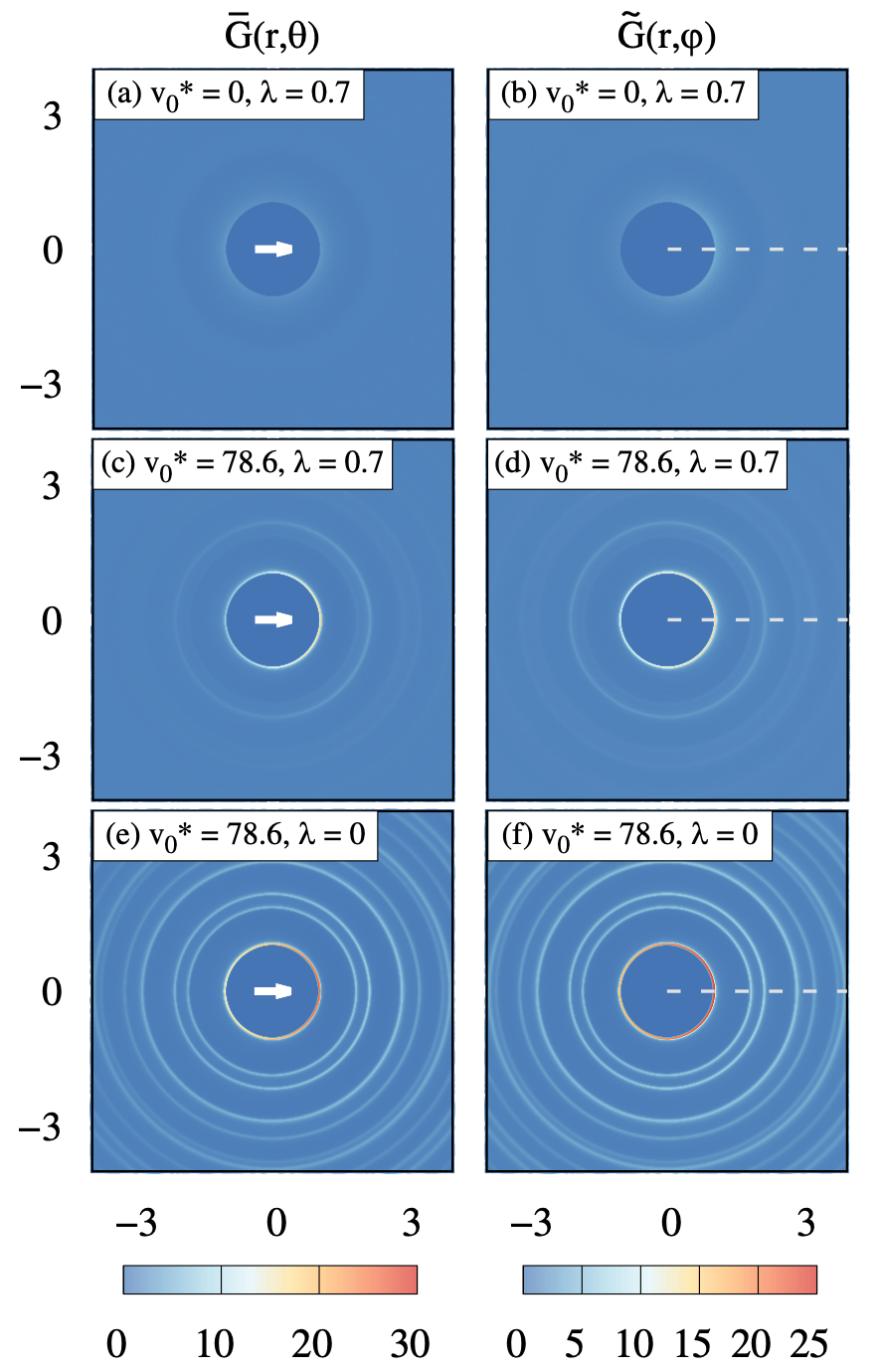}
   \caption{Left column: Integrated correlation function $\bar{G}(r,\theta)$ for a) $v_0^* = 0$ and $\lambda = 0.7$; c) $v_0^* = 78.6$ and $\lambda = 0.7$; e) $v_0^* = 78.6$ and $\lambda = 0$. The arrow denotes the orientation of the tagged particle, relative to which $\theta$ is measured. Right column: Integrated correlation function $\tilde{G}(r,\varphi)$ b) $v_0^* = 0$ and $\lambda = 0.7$; d) $v_0^* = 78.6$ and $\lambda = 0.7$; f)  $v_0^* = 78.6$ and $\lambda = 0$. The dashed line indicates the relative orientation $\varphi=0$, corresponding to parallel orientation of the two dipole moments.}
    \label{results_fig:3} 
\end{figure}

Numerical results for correlation functions at different values of $v_0^*$ and $\lambda$ are presented in \cref{results_fig:3}. To disentangle the role played by the different angular variables, we separately show and discuss the integrated correlation functions $\bar{G}(r,\theta)=\int d\varphi G(r,\theta, \varphi)$ and  $\tilde{G}(r,\varphi)=\int d\theta G(r, \theta,\varphi)$. The function $\bar{G}(r,\theta)$ measures the probability to find a particle with arbitrary orientation at a relative distance $r$ from the tagged particle in the direction $\theta$, relative to the tagged particle's orientation. The second correlation function, $\tilde{G}(r,\varphi)$, captures the probability to find a pair of particles at distance $r$ and with a relative alignment of their dipole vectors $\varphi$, regardless of their relative position in space.

%SABINE'S SUGGESTION
We start by considering the passive case, $v_0^{*}=0$, at $\lambda=0.7$ (upper panel of \cref{results_fig:3}).
Here, as seen from \cref{results_fig:3}(a), the function $\bar{G}(r,\theta)$ is nearly independent of $\theta$ and exhibits only a weak spatial structure beyond the first peak.
We note that the symmetry with respect to the particle's axis (i.e., the axis of the dipole moment), $\theta\rightarrow \theta+\pi$, is expected due to the symmetries of the interaction (see \cref{2-bodycorrelfunct}). In strongly coupled passive dipolar fluids ($\lambda\gg 1$), the correlations are enhanced along the axis of the dipole moment ($\theta= 0$, $\theta=\pi$) but weakened in the equatorial plane ($\theta=\pi/2$, $\theta=3\pi/2$), signalling head-to-tail ordering into chains. However, here we are considering $\lambda=0.7$ where this tendency is not very pronounced, yielding only weak dependence on $\theta$. Considering the function $\tilde{G}(r,\varphi)$ in the passive case (\cref{results_fig:3}(b)), we observe a slight preference of parallel orientation, i.e., small values of $\varphi$, rather than anti-parallel orientation ($\varphi\sim \pi$).

Upon "switching on" the motility towards values within the MIPS regime (specifically, $v_0^{*}=78.6$), both correlation functions change, as seen from the middle panel in \cref{results_fig:3}. In particular, the function $\bar{G}(r,\theta)$ now exhibits a symmetry-breaking with respect to the particle's "equatorial plane", that is, the correlation is more pronounced in front of the particle ($\theta\sim 0$) than behind ($\theta\sim\pi$). In other words, it is more probable to find a neighbouring particle in front than behind the tagged particle. This reflects the trapping mechanism which eventually leads to phase separation. Also, the range of correlations is somewhat increased relative to the passive case. Both effects are reminiscent of previous results for pure ABP and ABP with simpler (alignment) interactions, see \cite{Bialke2013,Sese-Sansa2021}.
Interestingly, however, the anisotropy effects are much weaker in the present, dipolar case. This may be seen from comparing \cref{results_fig:3}(c) to \cref{results_fig:3}(e), which shows the $\bar{G}(r,\theta)$ of a pure ABP system at the same density and $v_0^{*}=78.6$. Clearly, the pure ABP system is characterized by an even enhanced anisotropy and longer-ranged correlations. We understand the overall weakening of correlations as follows: dipolar particles tend to form head-to-tail clusters. This tendency acts against the motility-induced, asymmetric agglomeration of neighbours in front of a particle.

Similar observations emerge from analysing the function $\tilde{G}(r,\varphi)$. Upon switching on activity for the dipolar system (compare Figs. 6((d) and (b)), the main effect is a slight increase of the first peak. However, this occurs at essentially all relative orientations $\varphi$, with only weak  preference of $\varphi\sim 0$ (parallel orientation) at the small value of $\lambda$ considered. 
In contrast, the $\tilde{G}(r,\varphi)$ of the corresponding pure ABP system ($\lambda=0$) at $v_0^{*}=78.6$ is characterized by several pronounced peaks (\cref{results_fig:3}(f)). 
Interestingly, also for this system, neighboring particles tend to orient their heading vectors in the same direction. Thus, there is to some extent a "velocity alignment" without any explicit aligning interactions. 
To understand this surprising effect, we recall that the pure ABP system at the parameters considered is deep inside the MIPS state, which implies the formation of a macroscopic cluster. At the cluster's interface, one expects that particles are on average pointing along the density gradient, i. e., towards the
center of the cluster \cite{Fily2012,Fily2014,Solon2018,Lauersdorf2021}.
%, resulting in an orientational correlation captured by $\tilde{G}(r,\varphi)$ ({\textcolor{red}{we should discuss this}).
 
To summarize, we see that a finite motility does affect the two-body correlations of the (weakly coupled) dipolar system; in particular, motility induces an anisotropy (with respect to the distribution in front and behind a particle) not seen in the passive case.  
However, these effects are much less pronounced than in a corresponding pure ABP system. This reflects, at a microscopic level, the hindering of the particle trapping mechanisms and thus, of MIPS, by dipolar interactions.
 
Having calculated the full correlation function $G(r,\theta,\varphi)$, it is straightforward to obtain
numerical values for the translational friction coefficient $\tilde{\zeta}$ according to \cref{deriv_hydro-eq:12}. Due to the long-range character of the dipole-dipole interaction, some care is required when choosing the cut-off of the integration.
Here we set the cut-off to $5R$ which proves to be sufficient for the parameters considered.

\begin{figure}
\centering
\includegraphics[trim=30 70 30 70, width=\columnwidth]{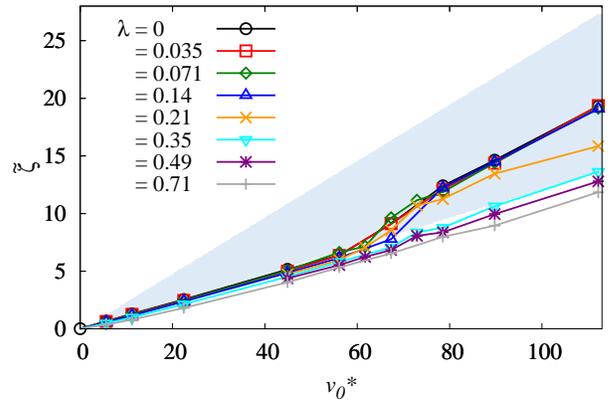}
\caption{Numerical values of the dimensionless translational friction coefficient $\tilde{\zeta}$ as a function of the motility, computed from Brownian dynamics simulations, for different values of the coupling strength $\lambda$. Also shown is the instability (blue) region predicted by the hydrodynamic theory (\cref{modeldescrip_eq:22}) at $\tilde{\varepsilon} = 0$.}
\label{mapping_theories_fig:1}
\end{figure}

Results for the functions $\tilde{\zeta}(v_0^{*})$ at different values of $\lambda$ are shown in \cref{mapping_theories_fig:1}.
For $v_0^{*}\rightarrow 0$, all functions approach zero, as expected in the passive case due to corresponding symmetries
of $G(r,\theta,\varphi)$ (see \cref{symmetries_correlat}). Upon increase of the motility from zero, the functions, $\tilde{\zeta}(v_0^{*})$ grow monotonically, indicating
an increase of "translational friction" due to the trapping mechanism (as reflected by $G$) discussed before.
However, the details depend on $\lambda$. Generally, the value of $\tilde{\zeta}$ at a given $v_0^{*}$ is the smaller, the larger $\lambda$.
This can be understood from our earlier analysis of the correlation functions, revealing that the motility-induced anisotropies (leading to non-zero friction) are hindered by dipolar interactions.
Moreover, the curves at small $\lambda$ exhibit a "kink" in the range $v_0^{*}\approx 60-70$. This is within the range of motilities where, according to \cref{results_fig:1.1}, MIPS occurs. In contrast, the functions $\tilde{\zeta}(v_0^{*})$ at larger coupling strengths do not exhibit a clear "kink". This might be related to the disappearance of MIPS  in the parameter range investigated.

Finally, we compare the behavior of these functions with the stability limit $\tilde{\zeta}_{\pm}(v_0^{*})$ predicted by the theoretical analysis in \cref{theoreticaldesc}, see \cref{modeldescrip_eq:22}.
For small values of the motility, the numerically obtained functions $\tilde{\zeta}(v_0^{*})$ lay outside the instability (blue) region for all values of $\lambda$ considered.
This can be interpreted such that the particle-resolved simulations, consistent with the theory, predict the homogeneous and isotropic state to be stable in this parameter regime. 
Upon increasing $v_0^{*}$, the values of $\tilde{\zeta}$ for the weakly coupled systems penetrate into the instability region right in the range of motilities where the numerical phase separation occurs.
Moreover, this happens the later, the larger $\lambda$, indicating that the destabilization is shifted to higher motilities when $\lambda$ increases. This shift is consistent with the numerical results in \cref{results_fig:1.1}, and it is a direct consequence of the hindering of trapping in the presence of dipolar interactions

All in all, the comparison of the numerically obtained translation friction coefficients $\tilde{\zeta}(v_0^{*})$ with the stability limit predicted by the (mean-field-like) hydrodynamic theory
shows quite consistent behavior. Note that we have focused here on particular values of the overall density and repulsion strength, that is, parameters which are important for the onset of MIPS already for pure ABP 
(as discussed in \cref{limits_continuum_theory}). Still we can conclude that, at small densities, the relatively simple hydrodynamic theory predicts MIPS in the dipolar active system quite well.

\section{Conclusions}
In this paper, we propose a hydrodynamic description of systems of dipolar ABP, starting from the microscopic dynamics. This bottom-up approach allows us to establish a direct link between the coarse-grained model and the over-damped Langevin dynamics. Interestingly, we find that, due to the symmetries in the pairwise correlations imposed by the dipolar interaction potential, the resulting torque enters the coarse-grained description in the same manner as it does for torques deriving from Vicsek-like alignment rules \cite{Sese-Sansa2021}. 
We study, up to linear order, the destabilization of the homogeneous and isotropic phase. We find that the destabilization mechanism is governed by a long wavelength instability, which we identify with MIPS. 

In parallel, we perform Brownian dynamics simulations of a system of dipolar ABP in the weak-coupling regime $(\lambda < 0.71)$ and show that dipole-dipole interactions suppress MIPS, as first reported in \cite{Liao2020} (yet at larger density). Analysing the angular correlation functions we find that the trapping mechanism familiar from systems of pure ABP is weakened in the presence of dipolar coupling. Indeed, dipolar interactions rather favour head-to-tail configurations, which act to oppose MIPS. 
%to the one describing systems of to . This highlights the generality of the continuum model.
%the torque enters in the same way
Finally, we exploit the direct mapping between the coarse-grained theory and particle-based simulations to obtain an alternative prediction of the onset of MIPS. This again confirms the hindering of MIPS in the presence of weak dipole-dipole interactions. Our approach can therefore be considered as a complementary tool to detect MIPS in complex active systems.
%We emphasise the predictive character of the model to locate the onset of phase separation. 

A possible next step would be to extend the present formalism to account for the breaking of rotational symmetry leading to a polar flocking state \cite{Liao2020}. In this case, however, one needs to modify certain steps of the derivation. In particular, one has to choose another frame of reference to define pairwise correlations, since the body-fixed frame used in the present study is no longer appropriate if the system is not globally isotropic. Based on such a modified theory, 
one would be able to explore the coupling between the flocking phase transition and MIPS in strongly coupled dipolar active systems, as well as in other active systems with different types of symmetries and order parameters.

\appendix
\section{Effective coefficients for force and torque: symmetries and integrated numerical values}\label{symmetries_correlat}

In this section we first consider in detail the properties of the correlation function $G(r_{12},\theta_1,\varphi_{12})$. We start with the passive dipolar case, for which the symmetries of the interaction potentials transfer to the correlation function. This paves the way to discuss the active case, in which the non-equilibrium nature of the system induces additional anisotropies. We then discuss consequences for the coefficients $\tilde{\zeta}$ and $\tilde{\varepsilon}$. 

\subsection{Two-body correlation function}\label{2-bodycorrelfunct}

\paragraph*{Passive case} 
In a passive system of dipolar particles, the angular properties of $G(r_{12},\theta_1,\varphi_{12})$ are determined by those of the (conservative) pair potential. Since the steric potential contains no angular dependencies, we here focus on the symmetries exhibited by the dipole-dipole potential, $u_{dd}$. 
Within our choice of variables, we have $u_{dd}\propto\left(\cos\varphi_{12}-3\cos\theta_1\cos \theta_2\right) = \left(\cos\varphi_{12}-3\cos\theta_1\cos(\varphi_{12}+\theta_1)\right)$. 
%According to the addition theorem, one has
%$\cos(\varphi_{12}+\theta_1)=\cos\varphi_{12}\cos\theta_1-\sin\varphi_{12}\sin\theta_1$. With this, the dipolar potential becomes
%$u_{DD}\propto\left(\cos\varphi_{12}-3\left(\cos^2\theta_1\cos \varphi_{12}-\cos\theta_1\sin\theta_1\sin\varphi_{12} \right)   \right)$.
We thus observe the following symmetries: 
\begin{enumerate}
\item $\varphi_1=\varphi_1+\pi$ and $\varphi_2=\varphi_2+\pi$, such that $\varphi_{12}=\varphi_2-\varphi_1$ remains constant. This transformation corresponds to a reversal of the dipole vector of both particles (i.e., $\boldsymbol{\mu}_i\rightarrow -\boldsymbol{\mu}_i$, $i=1,2$).
\item $\theta_1\rightarrow \theta_1+\pi$ (note that $\cos(\theta_1+\pi)=-\cos\theta_1$ and therefore, products of this function are invariant). Physically, this invariance expresses the fact that, in the passive case, the probability to find a second particle (irrespective of its orientation) in front or behind the tagged particle is the same. In other words, the integrated function $\bar{G}(r_{12},\theta_1)=\int d\varphi_{12} G(r_{12},\theta_1, \varphi_{12})$ exhibits the symmetry
$\theta_1\rightarrow \theta_1+\pi$.
\item The symmetry of $u_{dd}$ against exchange of the dipole moments of particle $\textit{1}$ and particle $\textit{2}$ implies an invariance against the transformation $\varphi_{12}\rightarrow -\varphi_{12}$, accompanied by $\theta_1 \rightarrow \theta_2$ and $\theta_2 \rightarrow \theta_1$. This is a consequence of the non-separable nature of $u_{dd}$. However, the integrated function $\tilde{G}(r_{12},\varphi_{12})=\int d\theta_1 G(r_{12}, \varphi_{12},\theta_1)$, which measures the relative alignment of the two particles irrespective of the direction of the connection vector, must fulfil $\varphi_{12}\rightarrow -\varphi_{12}$.
\item As argued in 3., the symmetry against the transformation $\theta_1\rightarrow -\theta_{1}$ alone is not fulfilled and must be accompanied by the transformation $\varphi_{12}\rightarrow -\varphi_{12}$ in order to leave $u_{dd}$ invariant. Nonetheless, if one considers the integrated function $\bar{G}(r_{12},\theta_1)$, which does not depend on the relative orientation between pairs of particles, then the symmetry $\theta_1\rightarrow -\theta_{1}$ is fulfilled.
\end{enumerate}

\paragraph*{Active case} At finite $v_0^*$, the symmetry 2. of the passive system breaks down.
%one would expect that symmetry a) is still maintained. However, symmetry b) is broken. Indeed, 
Indeed, as it is seen from our numerical results for the integrated correlation function $\bar{G}(r_{12},\theta_1)=\int d\varphi_{12} G(r_{12},\theta_1, \varphi_{12})$, \cref{results_fig:3}, the probability to find a second particle (irrespective of its orientation) is higher in front of a tagged particle, than behind it.
Mathematically, this means that $G(r_{12}, \theta_1,\varphi_{12})$ depends on $\theta_1$ via odd powers of $\cos\theta_1$ (or products of $\cos^n$ and $\sin^m$ such that $m+n$ is odd). However, symmetries 1., 3. and 4. are still maintained even in the active case.

\subsection{The $\tilde{\zeta}$ coefficient } \label{appxzetacoeff}

Inspecting the integrand in the expression for $\tilde{\zeta}$, \cref{deriv_hydro-eq:12}, and using the addition theorem (i.e., $\cos(\varphi_{12}+\theta_1)=\cos\varphi_{12}\cos\theta_1-\sin\varphi_{12}\sin\theta_1$), we find that all terms stemming from the derivative of the dipolar potential 
change their sign upon the transformation $\theta_1\rightarrow \theta_1+\pi$ (because they contain $\cos\theta_1$, $\cos^3\theta_1$, or $\cos^2\theta_1\sin\theta_1$). In the passive case, $G(r_{12}, \varphi_{12},\theta_1)$ is invariant against this transformation (see the symmetry argument b) above). Therefore, the integral $[0,2\pi]$ over $\theta_1$ vanishes, yielding $\tilde{\zeta}=0$.
In the active case, however, $G$ is not any more invariant against $\theta_1\rightarrow \theta_1+\pi$ (as argued above). Therefore, we obtain non-zero values for all terms entering the force coefficient.

\subsection{The $\tilde{\varepsilon}$ coefficient }\label{appxepscoeff}

As seen from \cref{deriv_hydro-eq:14}, the coefficient $\varepsilon$ contains two terms in the integral. The first term involves the product $\sin\varphi_{12}\times G(r_{12}, \theta_1,\varphi_{12})$. We can therefore first perform the integral $\int d\theta_1 G(r_{12}, \theta_1,\varphi_{12})$, yielding $\tilde{G}(r_{12},\varphi_{12})$. From our discussion in \cref{2-bodycorrelfunct}, $\tilde{G}(r_{12},\varphi_{12})$
exhibits a mirror symmetry $\tilde{G}(r_{12},\varphi_{12})=\tilde{G}(r_{12},-\varphi_{12})$, contrary to the function $\sin\varphi_{12}$ with which it is multiplied. Thus, the integral $\int_0^{2\pi}d\varphi_{12}$ vanishes. This holds both in the active and in the passive case.

The second term in the integral involves the product $\sin\theta_1\cos(\varphi_{12}+\theta_1)\times G(r_{12},\theta_1, \varphi_{12})$. As discussed in \cref{2-bodycorrelfunct}, the integrated function $\bar{G}(r_{12},\theta_1)$ exhibits the symmetry $\theta_1\rightarrow -\theta_{1}$ (i. e.  $\bar{G}(r_{12},\theta_1) =  \bar{G}(r_{12},-\theta_1)$) both in the active and in the passive case, contrary to the odd function $\sin\theta_1$ with which it is multiplied. Therefore, the integral over $\theta_1$ vanishes, leading to $\tilde{\varepsilon}=0$. As a result, the torque coefficient $\tilde{\varepsilon}$ is always identically 0.

\begin{figure}[ht]
\centering
\includegraphics[trim=50 50 0 70, width=\columnwidth]{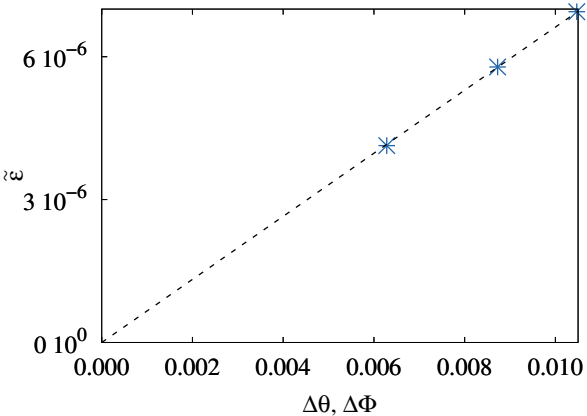}
\caption{Numerical values of the normalized rotational friction coefficient $\tilde{\varepsilon} = \frac{\bar{\rho}}{D_r}\varepsilon$ for a system at $v_0^*=112$ and $\lambda=0.71$, computed varying simultaneously the bin size $\Delta \theta $ and $\Delta \varphi $ of both the histogram and the discretization of the integrals. The dashed line is a fit of the data points and corresponds to the linear function $f(x)=0.000662x$.}
\label{appx_rotatcoeff_fig:1}
\end{figure}

To further convince ourselves that $\tilde{\varepsilon}$ indeed vanishes, we have numerically computed its value from \cref{deriv_hydro-eq:14}, with $G$ obtained from our simulations. Due to small discretization errors arising when computing $G(r_{12}, \theta_1,\varphi_{12})$ and when performing the integration,  $\tilde{\varepsilon}$ always has a finite, yet small, value. However, as shown in \cref{appx_rotatcoeff_fig:1}, a decrease of the bin size leads to smaller values of $\tilde{\varepsilon}$, which tend to 0 in the continuum limit.

\section{Technical details of the Brownian dynamics simulations}\label{browniandyndetails}

In the following, we first describe the parameter set employed to perform simulations. %Subsequently, we show 
We define the unit of length to be the cutoff distance of the WCA potential, $R$, and we set the strength of this potential to $u_0 =100 k_B T $. Thus, the thermal energy $k_B T=\beta^{-1} $ defines the energy unit, where $k_B$ is the Boltzmann constant and $T$ the temperature. Further, the time unit is given by $\tau = \frac{R^2}{D_t}$, where $D_t=1$. Particles are subjected to both translational and rotational noise. The translational and rotational diffusion coefficients fulfil the Stokes-Einstein relation, $D_r = 3 \frac{D_t}{R^2}$.
  We define the average packing fraction of the system as $\phi = \frac{N \pi R^2}{4L^2}$.
 Writing the equations of motion \cref{brown-dyn-mod_eq:1,brown-dyn-mod_eq:2} in dimensionless form, one can identify the relevant parameters to be the P\'eclet number (or motility) $v_0^* =v_0 \frac{ R}{D_t} $, which also determines the particles' persistence length, the WCA interaction strength $u_0$, and the dipolar interaction strength, $\lambda = \beta \mu^2 R^{-3}$, where $\mu$ is the magnitude of each dipole moment.
  
 We perform simulations of $N=1156$ particles in a square box of size $L \times L$ with periodic boundary conditions (PBC). The dimensionless timestep is set to $\delta t = 2^{-3} \times 10^{-5} \approx 8 \times 10^{-6}$. We employ an Euler-Maruyama scheme to integrate the equations of motion, \cref{brown-dyn-mod_eq:1,brown-dyn-mod_eq:2}. A two-dimensional Ewald summation is implemented to account for the long-range nature of the dipolar interactions \cite{Liao2020, Liao2020correction}. Indeed, even in 2D, simple truncation can lead to errors visible, e.g., in correlation functions \cite{mazars2011}. All simulations are initialized with randomly oriented particles being placed on a square lattice. The system typically reaches a steady state after $ 4 \times 10^{6}$ timesteps. Then the production run starts, generating a "snapshot" of the particles' configuration every 1000 timesteps, over which ensemble averages are performed to compute the quantities of interest.
Our goal is to analyse the impact of dipolar interactions on the onset of MIPS. To this end, we explore the $(\lambda,v_0^*)$ plane, keeping the global packing fraction fixed at $\phi = 0.4$. We emphasize the large computational cost of carrying out simulations where dipolar interactions are treated by means of an Ewald summation. As a result, some of our figures show only a reduced number of data points. This concerns, in particular, the state diagram in \cref{results_fig:1.1}.

\begin{figure*}
\centering
\includegraphics[trim=230 150 180 270 ,width=\columnwidth]{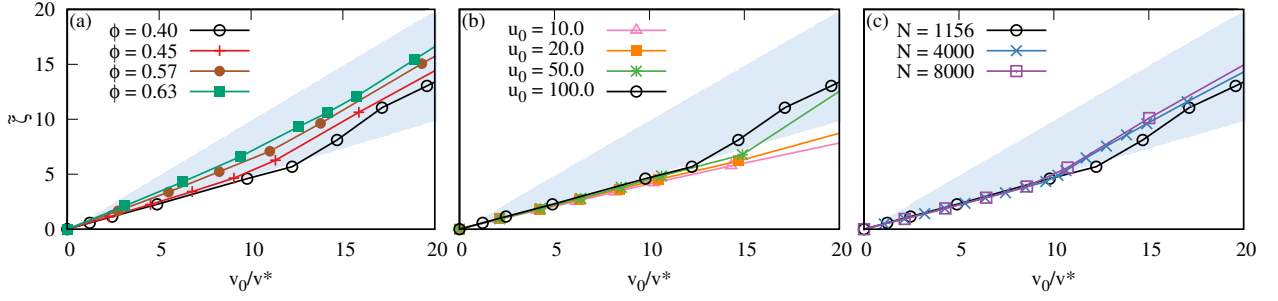}
\caption{Instability (blue) region at $\tilde{\varepsilon} = 0$ together with numerical values of the dimensionless translational friction coefficient $\tilde{\zeta}$ as a function of the dimensionless motility $v_0/v^*$. The different parameter sets correspond to suspensions of ABP at: (a) fixed repulsive interaction strength $u_0=100.0$, system size $N=1156$ and varying average packing fraction. (b) fixed average packing fraction $\phi=0.4$, system size $N=1156$ and varying repulsive interaction strength. (c) fixed average packing fraction $\phi=0.4$, strong repulsive interaction strength $u_0=100.0$ and varying system size.}
 \label{appx-limits_fig:1} 
\end{figure*}

\section{Stability predictions for a pure ABP system}\label{limits_continuum_theory}

In this section, we discuss the limitations of the linear stability analysis performed on \cref{deriv_hydro-eq:20,deriv_hydro-eq:21}. As argued in the main text, the onset of MIPS occurs when the numerical values of $\tilde{\zeta}$ enter into the instability region in the ($v_0/v^*,\tilde{\zeta}$) plane. Here, $v^*=4\sqrt{\Da D_r}$ and $\Da$ corresponds to the long-time diffusion coefficient of a system of passive disks ($v_0 = 0$). We consider a system of pure ABP and numerically compute $\tilde{\zeta}$ for different parameter sets, varying the mean area fraction, $\phi$, the strength of the repulsive WCA potential, $u_0$, and the system size, $N$. 

First, we observe from \cref{appx-limits_fig:1} (a) that the theoretical prediction worsens as $\phi$ grows: at $\phi \geq 0.45$, the numerical values of $\tilde{\zeta}$ lay in the unstable (blue) region even at low motilities, which does not allow for a quantitative prediction of the onset of MIPS. This is to some extent expected: the larger $\phi$, the more $\tilde{\zeta}$ depends on the actual correlation function. This dependency is neglected by the mean-field like approximation, resulting in an incorrect prediction of the instability (blue) region. 

The strength of the WCA potential is also a crucial parameter that could alter the performance of the theory, see \cref{appx-limits_fig:1} (b). We find that for softer particles (lower $u_0$), the numerical $\tilde{\zeta}$ always lays outside the instability region in the range of $v_0/v^*$ explored, see the curves for $u_0 =10, 20$ in \cref{appx-limits_fig:1} (b). Presumably, one should increase the self-propulsion velocity in order to see MIPS. However, we did not systematically investigate a broader range of $v_0/v^*$, but rather restricted ourselves to 'hard' repulsive particles $(u_0 =100.0)$, for which we know that MIPS takes place in the range of $v_0/v^*$ explored here. 

% coefficient, which quantifies the collision persistence, is not able to correctly capture the onset of phase separation, and the $\tilde{\zeta}$ points always lay outside the instability region in the range of $v_0/v^*$ explored, see the curves for $u_0 =10, 20$ in \cref{appx-limits_fig:1} (b).

Finally, our simulations indicate that finite size effects do not have a major impact, as evidenced in  \cref{appx-limits_fig:1} (c). The results show that, despite minor differences in the location of the onset of MIPS, the numerical values of $\tilde{\zeta}$ computed at different system sizes behave similarly: they lay in the stable (white) region at low motilities and enter the unstable region at a critical $v_0/v^*$.

These results shed light on the limits of application of the present type of continuum theory to study MIPS in systems of ABP.

\section{Radial distribution function}\label{appx_rad_dist_funct}

\begin{figure}[ht]
\centering
\includegraphics[width=\columnwidth]{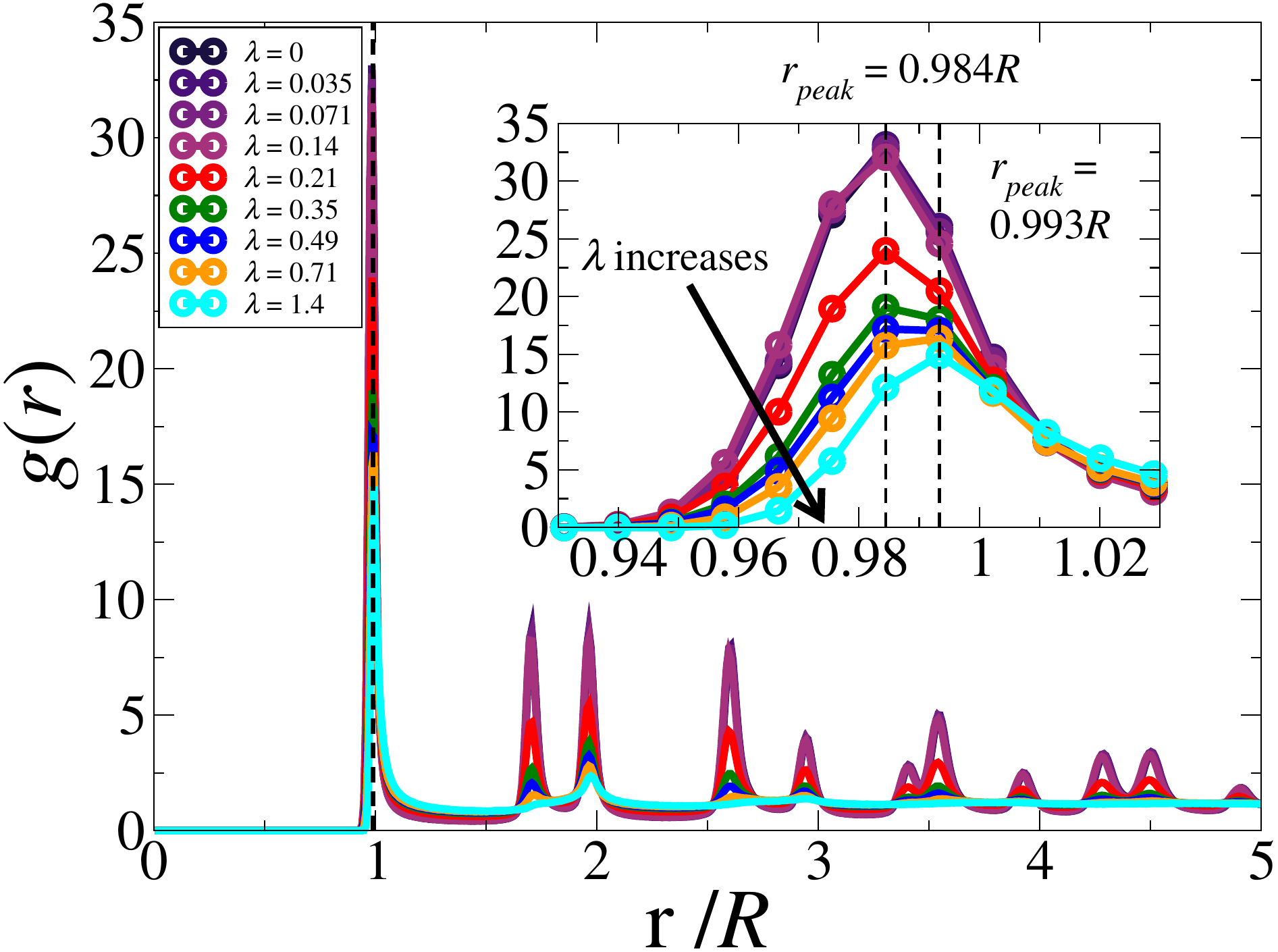}
\caption{Radial distribution function of a system of dipolar ABP at motility of $v_0^* = 112$, packing fraction $\phi=0.4$ and various dipolar coupling parameters, $\lambda$.}
\label{fig:1_appx} 
\end{figure}

To quantify the effect that dipolar interactions have on the characteristic distance between pairs of particles, we compute the radial distribution function $g(r)$ at fixed motility $v_0^*$ and varying the coupling parameter $\lambda$, \cref{fig:1_appx}.

We observe that an increase of the dipolar coupling shifts the first peak of the radial distribution function to somewhat higher distances, within our numerical accuracy. However, these distances are always smaller than the threshold distance imposed to define a cluster, $R$. Therefore, we do not need to adapt the cluster threshold distance to the value of $\lambda$ (in the range considered here).

\section*{Conflicts of interest}
There are no conflicts to declare.

\section*{Acknowledgements}
E.S.-S. and I.P. acknowledges Swiss National Science Foundation Project No. 200021-175719. D.L. acknowledges MCIU/AEI/FEDER for financial support under Grant Agreement No. RTI2018-099032-J-I00. I.P. acknowledges support from Ministerio de Ciencia, Innovaci\'on y Universidades MCIU/AEI/FEDER for financial support under grant agreement PGC2018-098373-B-100 AEI/FEDER-EU and from Generalitat de Catalunya under project 2017SGR-884. S.H.L.K. thanks the German Research Foundation for financial support via the projects 449485571 and 163436311-SFB 910.

%%%END OF MAIN TEXT%%%


%\balance

%\renewcommand\crefname{Notes and references}

%%%REFERENCES%%%
\bibliography{dipolarABPpreprint} %You need to replace "rsc" on this line with the name of your .bib file

\end{document}